\documentclass[aps,prb,twocolumn,superscriptaddress,showpacs]{revtex4-1}
\usepackage{amsmath,amssymb,graphicx}

\usepackage[T1]{fontenc}
\usepackage{chancery}

\usepackage{xcolor}
\usepackage{physics}
\usepackage[bottom]{footmisc}

\IfFileExists{newtxtext.sty}   {\usepackage{newtxtext,newtxmath}}
{\IfFileExists{stix.sty}
	{\usepackage{stix}}
	{\IfFileExists{mathptmx.sty}
		{\usepackage{mathptmx}}{} } }

\usepackage{textcomp}

\usepackage{bm}

\IfFileExists{siunitx.sty}{\usepackage{booktabs,siunitx}}{}

\usepackage{color}
\definecolor{LinkColor}{rgb}{0.256,0.439,0.588}
\usepackage{hyperref}
\hypersetup{
	pdfauthor={Tsung-Cheng Lu},
	pdftitle={Renyi Entropy of Chaotic Eigenstates},
	colorlinks=true,
	citecolor=LinkColor,
	linkcolor=LinkColor,
	urlcolor=LinkColor
}

\usepackage{pifont}
\usepackage{epstopdf}
\usepackage{caption}
\usepackage{subcaption}
\renewcommand{\raggedright}{\leftskip=0pt \rightskip=0pt plus 0cm}
\begin{document}
	
	\title{Renyi Entropy of Chaotic Eigenstates}

	\author{Tsung-Cheng Lu} 

	\author{Tarun Grover}

\affiliation{Department of Physics, University of California at San
		Diego, La Jolla, CA 92093, USA}
	\begin{abstract}
  Using arguments built on ergodicity, we derive an analytical expression for the Renyi entanglement entropies corresponding to the finite-energy density	eigenstates of chaotic many-body Hamiltonians. The expression is a universal function of the density of states and is valid even when the subsystem is a finite fraction of the total system - a regime in which the reduced density matrix is not thermal. We find that in the thermodynamic limit, only the von Neumann entropy density is independent of the subsystem to the total system ratio $V_A/V$, while the Renyi entropy densities depend non-linearly on $V_A/V$. Surprisingly, Renyi entropies $S_n$ for $n > 1$ are  convex functions of the subsystem size, with a volume law coefficient that depends on $V_A/V$, and exceeds that of a thermal mixed state at the same energy density. We provide two different arguments to support our results: the first one relies on a many-body version of Berry's formula for chaotic quantum mechanical systems, and is closely related to eigenstate thermalization hypothesis. The second argument relies on the assumption that for a fixed energy in a subsystem, all states in its complement allowed by the energy conservation are equally likely. We perform Exact Diagonalization study on quantum spin-chain Hamiltonians to test our analytical predictions, and find good agreement.

	\end{abstract}
	
%	\pacs{71.10.-w,71.10.Hf,75.40.Cx,75.40.Mg}
	
	\maketitle
	
%	\tableofcontents

\section{Introduction} \label{sec:intro}

The observation that the quantum evolution of a closed quantum system can lead to thermalization of local observables puts the foundations of equilibrium statistical mechanics on a firmer footing \cite{shankar1986,deutsch1991,srednicki1994chaos, srednicki1998,rigol2008,rigol_review,eisert_review}. In strong contrast to classical mechanics, where  one often refers to an ensemble of identically prepared systems, quantum mechanics allows for the possibility that a single quantum state can encode the full equilibrium probability distribution function, and in fact, the full quantum Hamiltonian \cite{garrison2015does}. Specifically, consider a system of size $V$ described by Hamiltonian $H$. The eigenstate thermalization hypothesis \cite{deutsch1991, srednicki1994chaos,srednicki1998} posits that the reduced density matrix  for a finite energy density eigenstate $|E_n\rangle$ on subsystem $A$ with $V_A \ll V$ is thermal: $\tr_{\overline{A}}|E_n \rangle \langle E_n| = \tr_{\overline{A}} \left( e^{- \beta H} \right)/ \tr \left( e^{- \beta H} \right) \stackrel{def}{=}
\rho^A_{\textrm{th}}(\beta) $ where $\beta^{-1} $ is the  temperature corresponding to the eigenstate $|E_n\rangle$ and equals $dS/dE\Big|_{E_n}$ where $S(E)$ is the microcanonical entropy (= logarithm of the density of states). In this work, we will employ the term `chaotic eigenstate' for the eigenstate $\ket{E_n}$ which obeys ETH.

One basic question is: do there exist observables $\mathcal{O}$ whose support $V_{\mathcal{O}}$ scales with the total system size $V$ while their expectation value  $\langle E_n| \mathcal{O}|E_n \rangle$ continues to satisfy some version of eigenstate thermalization? Standard analyses in statistical mechanics \cite{pathria_book} do not provide answer to such global aspects of thermalization. As pointed out in Ref.\cite{garrison2015does}, at any fixed, non-zero $V_A/V$, one can always find operators with operator  norm of order unity, for whom the difference $|\langle E_n | \mathcal{O}|E_n\rangle - \tr \left( \rho^A_{\textrm{th}}(\beta) O\right) |$ does not vanish and is  of order unity. This implies that the trace norm distance $\frac{1}{2} \bigg|\tr_{\overline{A}} |E_n \rangle \langle E_n| - \rho^A_{\textrm{th}}(\beta)\bigg|_1$ does not vanish and is of order unity when $V_A/V$ is held fixed while taking thermodynamic limit. Clearly, the expectation value of operators which are constrained by global conservation laws can't behave thermally. As an example, consider the operator $\left( H^2_A - \langle H^2_A \rangle \right)/V_A$, where $H_A$ is the Hamiltonian restricted to region A. Its expectation value in an eigenstate tends towards zero when $V_A$ approaches $V$, while is non-zero and proportional to the specific heat in a thermal state. This raises the question whether conserved quantities exhaust the set of operators that distinguish a pure state from a corresponding thermal state at the same energy?

\begin{figure}
	\centering
	\includegraphics[width=0.5\textwidth]{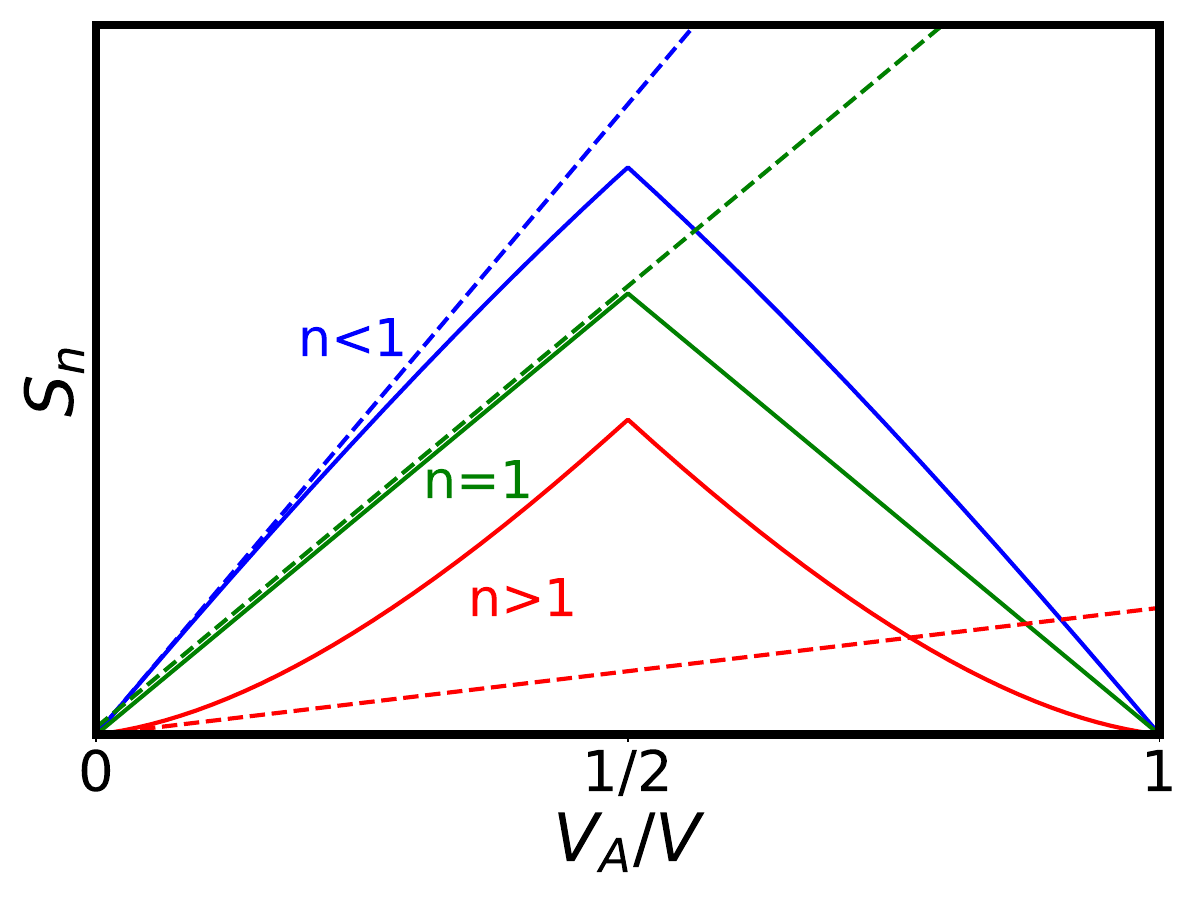}
	\caption{The curvature dependence of the Renyi entropy $S_n$ corresponding to  chaotic eigenstates derived in the main text (solid lines): in the thermodynamic limit, $S_n$ is a convex (concave) function of $ V_A/V$ for $n>1$ ($n<1$) with a cusp singularity at $V_A/V = 1/2$. The dashed lines correspond to the Renyi entropies of the thermal density matrix $\rho^A_{th}(\beta)=\exp(-\beta H_A)/Z$. $S_n$ of an chaotic eigenstate equals the thermal counterpart for $V_A/V<1/2$ at $n=1$, while for $n\neq 1$ equals the thermal counterpart only as $V_A/V\to 0$.} 
	\label{fig:sn_gaussian_compare_n_v3}
\end{figure}

One set of quantities that are particularly relevant to probe the global aspects of chaotic eigenstates are Renyi entropies: $S^A_n = \frac{1}{1-n}\log \left( \tr \,\rho_A^n\right)$. In fact $\tr \left( \rho^2 \right)$ is one of the simplest measures of how close  to a pure state a potentially mixed quantum state is. For integer values of $n$, $S^A_n $ has the interpretation of the expectation value of a cyclic permutation operator acting on the $n$ copies of the system. Due to this, $S^A_n$  can in principle be measured in experiments, and remarkably, an implementation for $n=2$ was recently demonstrated in cold atomic systems \cite{islam2015}.  

The ground states of quantum many-body systems typically follow an area-law for Renyi entropies (up to multiplicative logarithmic corrections): $S^A_n \sim L^{d-1}_A$ where $d$ is the spatial dimension \cite{bombelli1986,srednicki1993}. In strong contrast, finite energy density eigenstates of chaotic systems, owing to eigenstate thermalization, follow a volume law scaling: $S^A_n \sim L^{d}_A$ (see, e.g., \cite{deutsch2010}). Since we will often employ the term  `volume law coefficient', it is important to define it precisely. We define the volume law coefficient of an eigenstate as $\lim_{V \to \infty} S^A_n/V_A$ while keeping the ratio $V_A/V$ fixed and less than 1/2. Note that in principle this coefficient can depend on the ratio $V_A/V$ itself. For a thermal density matrix, $\rho = e^{-\beta H}/ \tr e^{- \beta H}$, the volume law coefficient is given by $n \beta (\mathfrak{f}(n \beta) -  \mathfrak{f}(\beta))/(n-1)$ where $\mathfrak{f}(\beta)$ is the free energy density at temperature $\beta^{-1}$. Therefore, in this example, the volume law coefficient is independent of $V_A/V$. Owing to eigenstate thermalization, the volume law coefficient of the Renyi entropy corresponding to chaotic eigenstates is also given by exactly the same expression, at least in the limit $V_A/V \rightarrow 0$. One of the basic questions that we will address in this paper is: what is the volume law coefficient corresponding to chaotic eigenstates when $V_A/V = O(1)$?

Ref. \cite{garrison2015does} provided numerical evidence that the the volume law coefficient for the von Neumann entropy $S^A_1$ corresponding to chaotic eigenstates equals its thermal counterpart even when the ratio $V_A/V\, (< 1/2)$ is of order unity. Furthermore, under the assumption that for a fixed set of quantum numbers in subsystem $A$, all allowed states in its complement $\overline{A}$ are equally likely, Ref.\cite{garrison2015does} provided an analytical expression for the n'th Renyi entropy $S_n$ for infinite temperature eigenstates of a system with particle number conservation. This expression curiously leads to the result that when $n \neq 1$, the Renyi entropies $S^A_n$ \textit{do not equal} their thermal counterpart for any fixed non-zero $V_A/V$ in the thermodynamic limit.

In a related development, Ref.\cite{dymarsky2016subsystem} also studied reduced density matrix corresponding to chaotic eigenstates of a system with only energy conservation. They found that the eigenvalues are proportional to the number of eigenstates of the rest of the system consistent with energy conservation. This result is very similar to the aforementioned result in Ref.\cite{garrison2015does} for the infinite temperature eigenstates with particle number conservation  - in that case the eigenvalues of the reduced density matrix were proportional to the number of eigenstates of the rest of the subsystem consistent with particle number conservation. Given this correspondence, one might expect that for  chaotic systems with only energy conservation,  only the von Neumann entropies equal their thermal counterpart, similar to the aforementioned example in Ref.\cite{garrison2015does}. This was already mentioned  in Ref.\cite{dymarsky2016subsystem} although Renyi entropies were not calculated.

In another development, Ref.\cite{fujita2017universality} studied `Canonical Thermal Pure Quantum states' (CTPQ) which were introduced in Ref.\cite{sugiura2013}. These states reproduce several features of a thermal ensemble while being a pure state \cite{sugiura2013}.  However, in contrast to the aforementioned result for infinite temperature eigenstates in Ref.\cite{garrison2015does},  the volume law coefficient of the Renyi entropies for CTPQ states is \textit{independent} of $V_A/V$ and equals the thermal Renyi entropy density. Ref.\cite{fujita2017universality} compared the Renyi entropy of eigenstates of non-integrable Hamiltonians with a fitting function based on CTPQ states.

In this paper, using a combination of arguments based on ergodicity and eigenstate thermalization, we derive an analytical expression for Renyi entropy of chaotic eigenstates. We follow two different arguments to arrive at the same result. Firstly, we consider a translationally invariant `classical' Hamiltonian $H_0$ (i.e. a Hamiltonian  all of whose eigenstates are product states) perturbed by an integrability breaking perturbation $H_1$ so that energy is the only conserved quantity for the full Hamiltonian $H = H_0 + \epsilon H_1$. Physical arguments and numerical results strongly suggest that if one first takes the thermodynamic limit, and only then takes  $\epsilon \rightarrow 0$, the eigenstates of $H$ are fully chaotic \cite{rigol2010quantum, flambaum1997criteria, georgeot2000quantum,santos2012chaos, rigol_srednicki2012, rigol2013, mukerjee2014,srednicki_kitp}. Following arguments inspired by Ref.\cite{srednicki1994chaos}, where eigenstates of a many-body chaotic system consisting of hard-sphere balls were studied, we argue that for ETH to hold for the  eigenstates of $H$, they may be approximated by random superposition of the eigenstates of $H_0$  in an energy window of order $\epsilon \ll V$. This can be thought of as a many-body version of the Berry's conjecture for chaotic billiard ball system where the eigenstates are given by random superposition of plane waves \cite{berry1977regular, srednicki1994chaos}. We will use the moniker ``many-body Berry'' (MBB) for such states. Related ideas have already been discussed in the context of one-dimensional integrable systems perturbed by a small integrability breaking term \cite{santos2012chaos, rigol_srednicki2012, rigol2013, rigol_asymmetry2016}.

In the second approach, we consider states of the form $\ket{\psi} =  \sum_{   E_i^A+E_j^{\overline{A}}  \in (E-\frac{1}{2}\Delta , E+\frac{1}{2} \Delta  )     }  C_{ij} \ket{ E_{i}^A} \otimes \ket{E_{j}^{\overline{A}}  },$ with $C_{ij}$ a random complex number, $\ket{ E_{i}^A}$ an eigenstate of $H_A$ and   $\ket{E_{j}^{\overline{A}}}$ that of $H_{\overline{A}}$. These states are exactly of the form suggested by `canonical typicality' arguments \cite{goldstein2006,popescu2006entanglement} and in the thermodynamic limit, reproduce the results of Ref.\cite{dymarsky2016subsystem} for the matrix elements of the reduced density matrix. Given the results in Ref.\cite{dymarsky2016subsystem}, it is very natural to conjecture that eigenstates of local Hamiltonians mimic states drawn from such an ensemble. We will call this  ``ergodic bipartition'' conjecture. The advantage of working with wavefunctions, in contrast to the average matrix elements of the reduced density matrix is that it allows us to calculate average of the Renyi entropy itself, which is a much more physical quantity compared to the Renyi entropy of the averaged reduced density matrix. This distinction is particularly crucial in finite sized systems. We will compare our analytical predictions with the exact diagonalization, as well as directly with the CTPQ states.

We first provide numerical evidence for both the `many-body Berry' conjecture as well as the `ergodic bipartition' conjecture by studying chaotic Hamiltonians of spin-chains. Next, we analytically calculate the Renyi entropies for such states. The analytical form of the results is identical in  either case. 

\vspace{0.3cm}

\underline{Our main results are:}

\begin{enumerate}
	\item Renyi entropies are a universal function of the density of states of the system. 
	
	\item Renyi entropy density $S^A_n/V_A$ depends on $V_A/V$ when $n \neq 1 $ as thermodynamic limit is taken. For $n > 1 (n<1) $, $S^A_n$ is always a convex (concave) function of $V_A/V$. $n=1$ corresponds to a transition point between concavity and convexity, and correspondingly the von Neumann entropy is linear in $V_A$ (see Fig.\ref{fig:sn_gaussian_compare_n_v3}). Consequently, in the thermodynamic limit for any non-zero $V_A/V$ , the volume law coefficient of the Renyi entropy $S^A_n$ differs from the one derived from the thermal density matrix $\rho^A_{th}(\beta)$ or equivalently the canonical thermal pure quantum state (CTPQ) states. For $n>1$, it exceeds that of a thermal/CTPQ  state, and for $n <1$, it is less than that of a thermal/CTPQ state.
	
	\item The Renyi entropy for a given $V_A/V$ depends on the density of states at an energy density that is itself a function of $V_A/V$. This allows one to obtain information about the full spectrum of the Hamiltonian by keeping the Renyi index $n$ fixed and only varying the ratio $V_A/V$. This is in strong contrast to the limit $V_A/V \rightarrow 0$ where $S^A_n$ only encodes thermodynamical information at temperature $\beta^{-1}$ and $(n \beta)^{-1}$.
\end{enumerate}

The paper is organized as follows: In Sec.II we state and provide evidence for the aforementioned many-body Berry conjecture and the ergodic bipartition conjecture for spin-chain Hamiltonians. In Sec III, we provide analytical results on Renyi entropies for the corresponding states, and discuss the salient features of our results. In particular, we discuss the curvature dependence of the Renyi entropies, as well as provide simple examples where one can obtain closed form expressions. In Sec. IV, we numerically study Renyi entropies corresponding to the spin-chain Hamiltonians and show that the results match our analytical predictions rather well. In Sec. V, we discuss the implications of our results, and future directions.

\section{The nature of Chaotic Eigenstates} \label{sec:eigen}

Consider a many-body Hamiltonian $H$ which we write as

\begin{equation}
H=H_A+H_{\overline{A}}+H_{A\overline{A}}, 
\end{equation}
where $H_A$,  $H_{\overline{A}}$ denote the part of $H$ with support only in real-space regions $A$ and $\overline{A}$ respectively, and $H_{A\overline{A}}$ denotes the interaction between $A$ and $\overline{A}$.  `Canonical typicality' arguments \cite{goldstein2006,popescu2006entanglement} imply that a typical state in the Hilbert space with energy $E$ with respect to $H$ has a reduced density matrix $\rho_A$ on region $A$ with matrix elements:

\begin{equation}\label{eq:rho_av}
\bra{E_i^A} \rho_A \ket{E_i^A}=\frac{1}{N} e^{S_{\overline{A}}(E-E_i^A)},
\end{equation}
where $\ket{E_i^A}$ is an eigenstate of $H_A$ with energy $E_i^A$, $e^{S_{\overline{A}}(E-E_i^A)}$ is the number of eigenstates of $H_ {\overline{A}}$ with energy $E_{\overline{A}}$ such that $E_A+E_{\overline{A}} \in (E-\frac{1}{2} \Delta , E+\frac{1}{2} \Delta )$ with $\Delta \ll E$ , and  $N$ is the total number of states in the energy window:

\begin{equation}
N=\sum_i e^{S_{\overline{A}}(E-E_i^A)} =\sum_{E_A} e^{S_A(E_A)+S_{\overline{A}}(E-E_A)} .
\end{equation}

One can obtain this result from  two conceptually different viewpoints. On the one hand, one can consider the following mixed state $\Omega$ that defines a microcanonical ensemble at energy $E$: 

\begin{equation}\label{eq:mixed}
\Omega=\frac{1}{N}  \sum_{   E_i^A+E_j^{\overline{A}}  \in (E-\frac{1}{2}\Delta , E+\frac{1}{2} \Delta  )     }  \ket{ E_{i}^A}\otimes  \ket{E_{j}^{\overline{A}}  } \bra{E_{i}^A} \otimes \bra{E_{j}^{\overline{A}} } ,
\end{equation}
\noindent
and then trace out the Hilbert space in region $\overline{A}$, thus obtaining Eq.\ref{eq:rho_av}. Alternatively, one can consider the following \textit{pure state} introduced in Refs.\cite{goldstein2006,popescu2006entanglement}:

\begin{equation}\label{eq:ergodic}
\ket{E} =  \sum_{   E_i^A+E_j^{\overline{A}}  \in (E-\frac{1}{2}\Delta , E+\frac{1}{2} \Delta  )     }  C_{ij} \ket{ E_{i}^A} \otimes \ket{E_{j}^{\overline{A}}  },
\end{equation}
where $C_{ij}$  is a complex random variable. After averaging, one again obtains Eq.\ref{eq:rho_av} when $V_A/V < 1/2$. The state in Eq.\ref{eq:ergodic} is the superposition of random tensor product of eigenstates of $H_A$ and $H_{\overline{A}}$ with the constraint of energy conservation, and we call it an ``ergodic bipartition'' (EB) state.

Recently, evidence was provided in Ref.\cite{dymarsky2016subsystem} that the reduced density matrix corresponding to an \textit{eigenstate} of translationally invariant non-integrable Hamiltonians resembles the reduced density matrix of a pure state based on canonical typicality, and therefore also satisfy Eq.\ref{eq:rho_av}.  
Therefore it is worthwhile to explore whether the state in Eq.\ref{eq:ergodic},  which leads to Eq.\ref{eq:rho_av}, is a good representative of the eigenstate of a chaotic Hamiltonian.

To explore this question, we first note that the state in Eq.\ref{eq:ergodic}  recovers the correct energy fluctuation in  an eigenstate \cite{garrison2015does}, namely, $\Delta E^2_A  = c T^2 \frac{V_A V_{\overline{A}}}{V_A + V_{\overline{A}}}$ (see Appendix \ref{sec:energyfluct}) where $c$ is the specific heat. Further, one  readily verifies that the diagonal entropy for a subsystem $A$ corresponding to this state equals the thermodynamic entropy $V_A s(E/V)$ where $s(x)$ denotes the entropy density at energy density $x$, as also expected from general, thermodynamical considerations \cite{polkovnikov_2011}.

Next, let's first see whether the ergodic bipartition states in Eq.\ref{eq:ergodic} satisfy ETH assuming that the eigenstates of $H_A$ and $H_{\overline{A}}$ are chaotic. Clearly if an operator is localized only in $A$ or $\overline{A}$, then its expectation value with respect to $\ket{E}$ trivially satisfies ETH by the very assumption that $H_A$ and $H_{\overline{A}}$ are chaotic. Therefore,  consider instead an operator $O = O_A \, O_{\overline{A}}$ where $O_A \in A$ and $ O_{\overline{A}} \in \overline{A}$. Recall that the ETH implies that $\bra{E_n} O \ket{E_m} = \overline{O(E/V)} \delta_{n,m} + \sqrt{\overline{O^2(E/V)}}e^{-S(\overline{E})/2} z_{n,m}$ where  $\overline{O(E/V)}$ is the microcanonical expectation value of $O$ at energy density $E/V$ and therefore is a smooth function of $E$, $S(\overline{E})$ is the microcanonical entropy at energy $\overline{E} = (E_n + E_m)/2$, and $z_{n,m}$ is a complex random number with zero mean and unit variance.

The diagonal matrix element of $O$ with respect to the state $\ket{E}$ in Eq. \ref{eq:ergodic} is given by:

\begin{eqnarray}
& & \bra{E} O \ket{E}  \nonumber \\
& & = \sum_{ij} |C_{ij}|^2 \bra{E^A_i} O_A \ket{E^A_i} \bra{E^{\overline{A}}_j} O_{\overline{A}} \ket{E^{\overline{A}}_j} 
\delta\left(E^A_i + E^{\overline{A}}_j-E\right) \nonumber \\
&&  =\sum_{E_A} \frac{e^{V_A s(E_A/V_A) + V_{\overline{A}} s((E-E_A)/V_{\overline{A}}) }  }{      e^{  Vs(E/V)  }} \overline{O_A(E_A/V_A)}\, \overline{ O_{\overline{A}}((E-E_A)/V_{\overline{A}})} \nonumber \\
&& = \overline{O_A(E/V)}  \,\,\overline{ O_{\overline{A}}(E/V)}
\end{eqnarray}
where the last equation in the sequence is derived by taking the saddle point from the one above. Clearly if $O_A$ and $ O_{\overline{A}}$ are located close to the boundary between $A$ and $\overline{A}$ (in units of thermal correlation length), then there is no reason to expect that $ \overline{O_A(E/V)}  \,\,\overline{ O_{\overline{A}}(E/V)}$ is the correct answer for the expectation of $O$ with respect to an actual eigenstate of the system. However, if $O_A$ and $ O_{\overline{A}}$ are located far from the boundary, then the cluster decomposition of correlation functions implies that the above answer is indeed correct to a good approximation. Note that it is a smooth function of the energy, as required by ETH. A similar calculation shows that the off-diagonal matrix element  $\bra{E_n} O \ket{E_m}$ is proportional to $e^{-S(\overline{E})/2} z$ where $\overline{E} = (E_n + E_m)/2$ and $z$ is a random complex number with zero mean and unit variance.

Above considerations indicate that the state $\ket{E}$ is a good representative of an eigenstate of $H$, \textit{except} for the correlation functions of operators close to the boundary. Therefore, we expect that it correctly captures the bulk quantities, such as the volume law coefficient of Renyi entropies. As already noted, it correctly reproduces the energy fluctuations, as well as the diagonal entropy for an eigenstate. Conversely, we do not expect it to necessarily reproduce the subleading area-law corrections to the Renyi entropies, which may be sensitive to the precise way the eigenstates of $H_A$ and $H_{\overline{A}}$ are `glued'.

In passing we note that Ref.\cite{deutsch2010} considered a perturbative treatment of the Hamiltonian $H = H_A + H_{\overline{A}} + \epsilon H_{A \overline{A}}$ to the first order in $\epsilon$. The wavefunctions thus argued to be obtained have  some resemblance with the EB state (Eq.\ref{eq:ergodic}). However, to really obtain an EB state via this procedure, one would instead need to  carry out the perturbation theory to an order that scales with the system size! This is because when $V_A/V$ is non-zero, the EB state has extensive fluctuations of energy in subregion $A$, unlike the states considered in Ref.\cite{deutsch2010} which essentially have no fluctuations since they mix eigenstates of $H_A$ in a small energy window.

\begin{figure}
	\centering
	\includegraphics[width=0.45\textwidth]{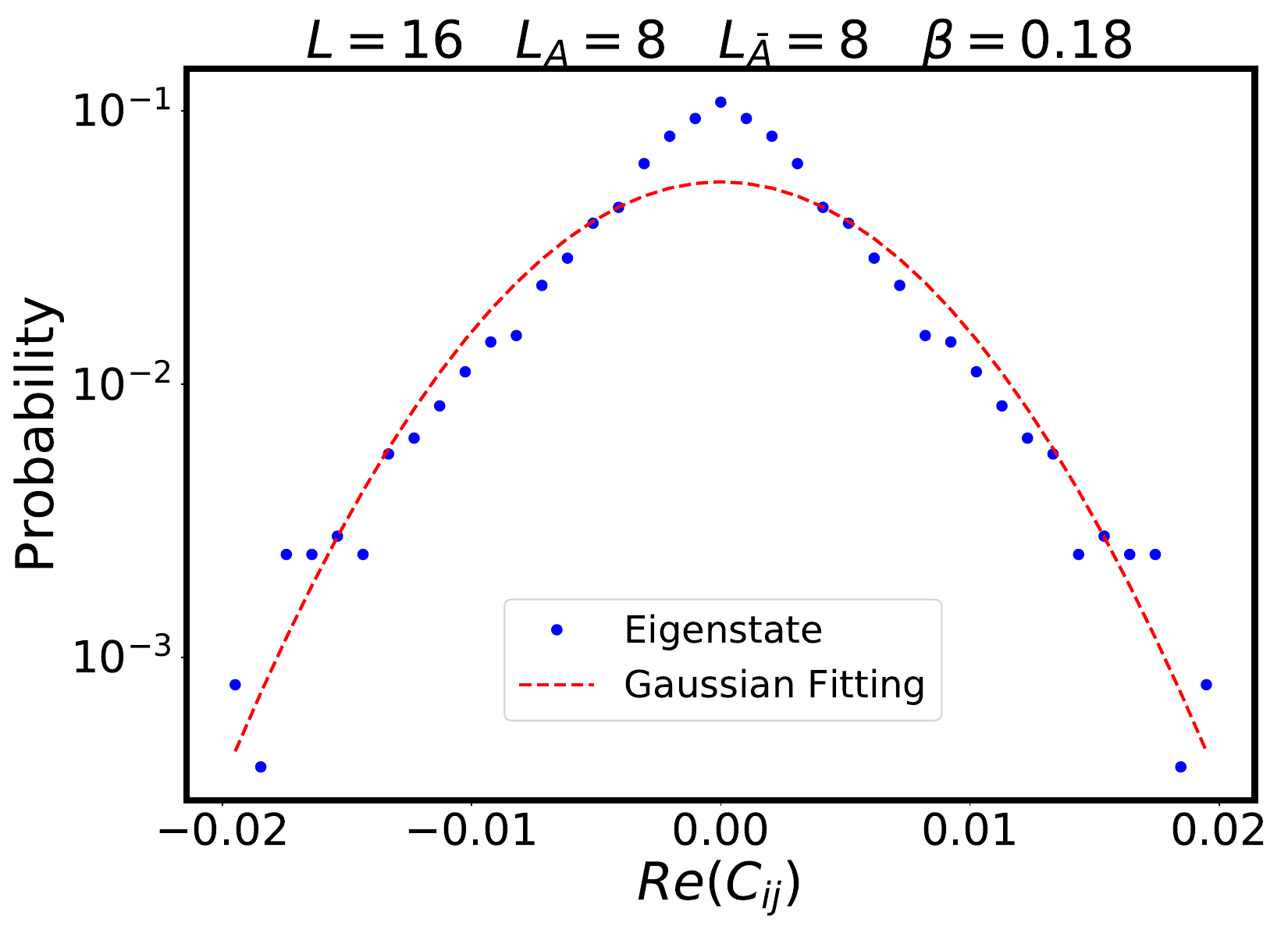}
	\caption{Probability distribution of the bipartite amplitudes $C_{ij}$ (Eq.\ref{eq:ergodic}) when $\ket{E}$ corresponds to a single eigenstate of the Hamiltonian in Eq.\ref{eq:spin_model}. The energy window $\Delta$ that appears in Eq.\ref{eq:ergodic} is chosen to be 2. The Gaussian distribution is obtained by least square fitting.} 
	\label{fig:amplitude_bipart}
\end{figure}

As a  numerical test of Eq.\ref{eq:ergodic}, consider a one dimensional spin-$1/2$ chain with the Hamiltonian given by 

\begin{equation}\label{eq:spin_model}
H=\sum_i^L -Z_iZ_{i+1}-Z_i +X_i,
\end{equation}
where we impose the periodic boundary condition $i\equiv i+L$.  Several works have already provided evidence in support of the validity of ETH in this model \cite{kim2014testing,banuls2015, zhang_huse2015, hosur2016,garrison2015does,dymarsky2016subsystem}.  By diagonalizing $H$, we calculate the bipartite amplitude of eigenstates on the bases of tensor product of all eigenstates of $H_A$ and $H_{\overline{A}}$ with $A$ denotes the sites $i=1,2,\cdots, L_A$ and $\overline{A}$ denotes the sites $i=L_A+1,L_A+2,\cdots, L$. Fig.\ref{fig:amplitude_bipart} shows the probability distribution of the bipartite amplitude on a semi-log plot. We find  deviations from a  Gaussian distribution. Although we do not understand the origin of this deviation, they may be due to the surface term unaccounted for in the definition EB states (\ref{eq:ergodic}). Nevertheless, as later shown, we find reasonable agreement for the Renyi entropies obtained from the EB state when compared to the exact diagonalization data.

A different starting point to obtain states that mimic chaotic eigenstates is provided by considering Hamiltonians $H$ of the form:

\begin{equation}\label{eq:MBB_H}
H=H_0+\epsilon H_1
\end{equation}
Here $H_0$ denotes a translationally invariant many-body local Hamiltonian whose eigenstates can be chosen as unentangled product states $\{\ket{s_\alpha} = \ket{s_1^ {\alpha}}\otimes\ket{s_2^{\alpha}}\otimes...\otimes\ket{s_L^{\alpha}}\}$, and therefore corresponds to an integrable system with an infinite number of conserved quantities. The $H_1$ term breaks the integrability. Physical arguments as well as numerics strongly suggest that when $H_1$ is local, the system will show a cross-over behavior from an integrable regime to a chaotic regime for $\epsilon \sim 1/L^{\beta}$
 \cite{rigol2010quantum, flambaum1997criteria, georgeot2000quantum,santos2012chaos,mukerjee2014,srednicki_kitp}. In fact, following arguments similar to Ref.\cite{srednicki1994chaos}, where eigenstates of a hard sphere system were written as random superposition of many-body plane waves so as to be consistent with ETH, in our case an eigenstate $\ket{E}$ of $H$ in the limit $\epsilon \rightarrow 0$ takes the form:

\begin{equation}\label{eq:random_state}
\lim_{\epsilon\to 0}\lim_{V \to \infty}  \ket{E} =\sum\limits_{\alpha} C_{\alpha} \ket{s_{\alpha}}
\end{equation}
with
\begin{equation}
P(\{ C_{\alpha}\})\propto \delta (1-\sum_{\alpha} |C_{\alpha}|^2 )\delta(E_{\alpha}-E),
\end{equation}
\noindent
where the first and second delta function constraints impose the normalization and energy conservation respectively.
This form of eigenstates closely resembles the Berry's conjecture for the  eigenstates of chaotic billiard ball systems \cite{berry1977regular}, and we will call this ansatz ``many-body Berry'' (MBB) conjecture. Again, similar to the case of ergodic bipartition conjecture discussed above (Eq.\ref{eq:ergodic}), one can readily verify that ETH holds true for the state in Eq.\ref{eq:random_state}. Specifically, the diagonal matrix elements of an operator $O$ match the canonical expectation value of $O$ with respect to $H_0$, while the off-diagonal matrix elements are proportional to $e^{-S(E)/2} z$ where $z$ is a random complex number with zero mean. Note that we take $H_0$ to be translationally invariant to avoid the possibility of many-body localization \cite{imbrie2016}.

\begin{figure}
	\centering
	\includegraphics[width=0.5\textwidth]{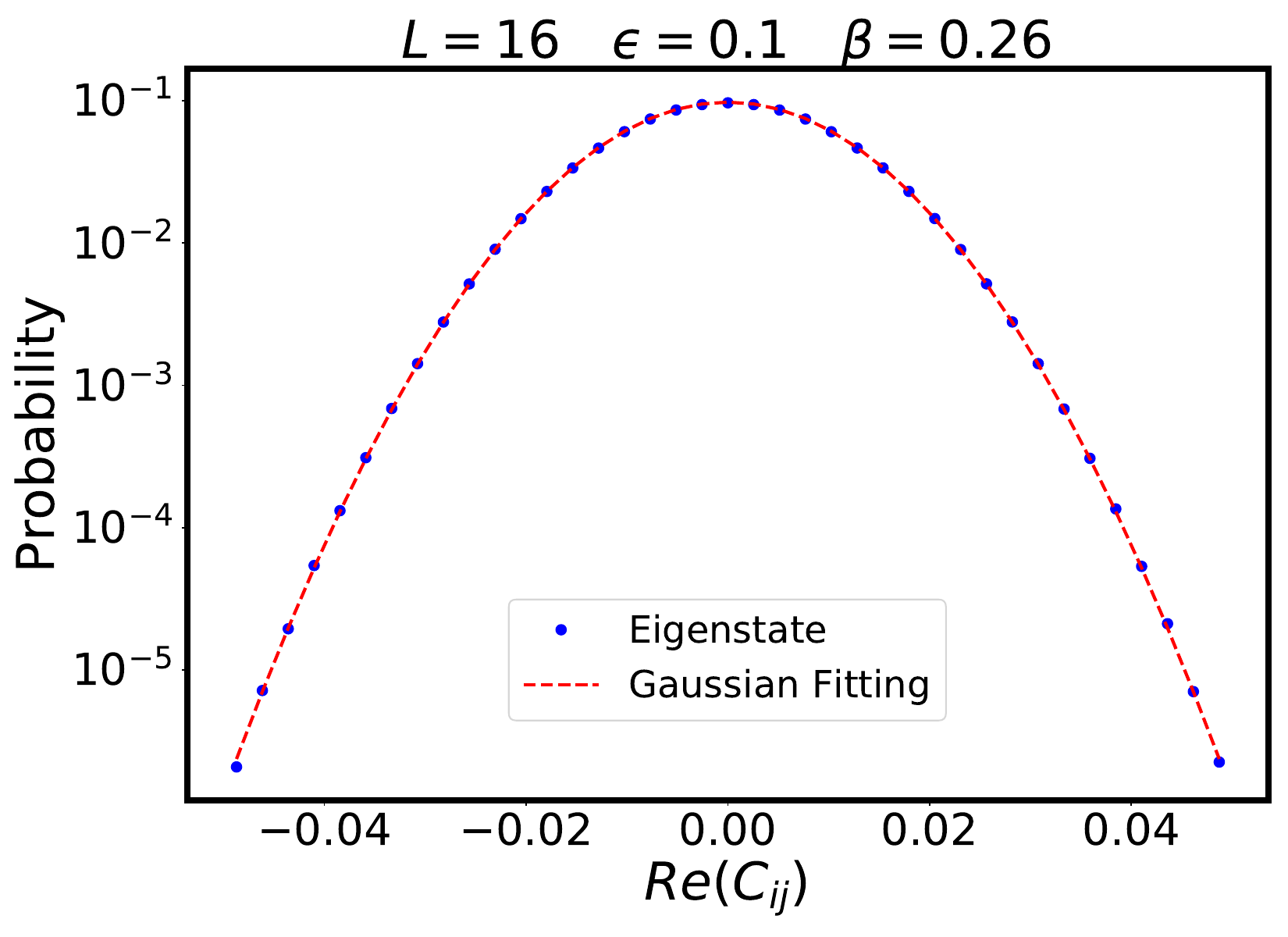}
	\caption{Probability distribution of the amplitudes $C_{\alpha}$ (Eq.\ref{eq:random_state}) for a single eigenstate of the Hamiltonian $H=-\sum_{i=1}^N Z_i  +\epsilon H_1$, where $H_1$ is a real hermitian random matrix. The Gaussian distribution is obtained by least square fitting.} 
	\label{fig:random_perturb}
\end{figure}

A quick demonstration of this conjecture is provided by the  Hamiltonian $H=-\sum_{i=1}^N Z_i  +\epsilon H_1$, where $H_1$ is a real hermitian random matrix. The  variance of the probability distribution function of the matrix element in $H_1$ is chosen such that the range of energy spectrum of $H_1$ is $L$. As shown in Fig.\ref{fig:random_perturb}, the coefficients $C_\alpha$ indeed behave as random Gaussian variables. Furthermore, we verified that their  variance equals $e^{-S}$, consistent with ETH. As we will discuss in detail in Sec. IV, one can consider a local perturbation, but the finite size effects are significantly larger with a local perturbation (i.e. the $\epsilon$ required to see chaos is comparatively larger), making it difficult to compare the eigenstates of $H$ with randomly superposed eigenstates of $H_0$. We again emphasize that all equal time correlation functions of the many-body Berry state (Eq.\ref{eq:random_state}) are determined fully by the properties of the Hamiltonian $H_0$ - the role of perturbation $H_1$ is `merely' to generate chaos.

\vspace{0.3cm}

%\newpage 
\vspace{1cm}

\textbf{Relation between Ergodic Bipartition States and Many-body Berry States:}

\vspace{0.2cm}

The many-body Berry states can essentially be thought of as a special case of ergodic bipartition states 
: if in Eq.\ref{eq:ergodic}, one substitutes for $\ket{ E_{i}^A}$ and $ 
\ket{E_{j}^{\overline{A}}}$ the eigenstates of $H_{0,A}$ and $H_{0,\overline{A}}$ respectively, where $H_{0,A}$ and $H_{0,\overline{A}}$ are restrictions of  the integrable Hamiltonian $H_0$ 
in Eq.\ref{eq:MBB_H} to region $A$ and $\overline{A}$, then the resulting state essentially corresponds to the many-body Berry state (Eq.\ref{eq:random_state}). However, there is a subtle distinction: the many-body Berry state does not suffer from any boundary effects due to the $H_{A \overline{A}}$ term: the  states $\ket{s_{\alpha}}$ that enter the definition of many-body Berry state in Eq.\ref{eq:random_state} are eigenstates of the Hamiltonian $H_0$ defined on the entire system. In contrast, the ergodic bipartition states involve tensor products of the  eigenstates of $H_{A}$ and $H_{\overline{A}}$, and therefore do not reproduce the correlations near the boundary between $A$ and $\overline{A}$ correctly, as discussed above.
Relatedly, comparing Fig.\ref{fig:amplitude_bipart} (ergodic bipartition conjecture), and Fig.\ref{fig:random_perturb} (many-body Berry conjecture), we notice that the latter figure fits the predicted Gaussian distribution better than the former. This is  likely again related to the  limitation of ergodic bipartition states, Eq.\ref{eq:ergodic}, that they suffer from boundary effects. Since we will concern ourselves only with the volume law coefficient of the Renyi entropies, we do not expect such boundary effects to be relevant.

Due to this relation between the ergodic bipartition states and the many-body Berry states, it turns out that from a technical standpoint, the calculations of their Renyi entropies - the central topic of our paper - are identical. This is the subject of our next section.

\section{Renyi Entropy of Chaotic Eigenstates} \label{sec:universalformula}

In this section we calculate Renyi entropy corresponding to the pure states in Eq.\ref{eq:ergodic} and Eq.\ref{eq:random_state}. We will not write separate equations for these two set of states, because as already mentioned, the calculation as well as all the results derived in this section apply to either of them. We will be particularly interested in the functional dependence of Renyi entropies on the ratio $V_A/V$. 

\subsection{Universal Dependence of Renyi Entropy on Many-body Density of States}  \label{sec:renyi_dos}

In principle, one can define three different kinds of averages to obtain Renyi entropies: (a) $S^A_n(\overline{ \rho_A}) =  \frac{1}{1-n} \log \left( \tr \left((\overline{\rho_A}) ^n\right)\right)$ (b) $S^A_n(\overline{\tr \rho^n_A}) = \frac{1}{1-n} \log\left( \overline{\tr \rho^n_A}\right)$ (c)  $S^A_{n,\textrm{avg}} =  \frac{1}{1-n} \overline{\log \left( \tr \left(\rho^n_A \right)  \right)}$. The physically most relevant measure is $S_{n,\textrm{avg}}$, however, it is also the hardest one to calculate due to averaging over logarithm.  As shown in Appendix \ref{sec:equiv}, the difference $|S_{n,\textrm{avg}} - S^A_n(\overline{\tr \rho^n_A}) |$ is exponentially small in the  volume of the total system. Due to this result and the fact that $S^A_n(\overline{\tr \rho^n_A})$ \textit{is} calculable using standard tools, in this paper we will focus mainly  on it, and with a slight abuse of notation, denote it as $\overline{S_n}$.  

One may still wonder how good is the measure (a), i.e., $S^A_n(\overline{ \rho_A})$, since it's the simplest one to calculate. Following Ref.\cite{popescu2006entanglement}, Levy's lemma implies that the trace norm distance between the average density matrix $\overline{\rho_A}$, and a typical density matrix of the ensemble vanishes exponentially in the total volume of the system. Combining this result with Fannes' inequality \cite{fannes1973}, $|S_1(\rho) - S_1(\sigma)| < |\rho - \sigma|_1 \log(\mathcal{H})$ where $\mathcal{H}$ is the size of the Hilbert space, one finds that in the thermodynamic limit, at least the von Neumann entropy for $\overline{ \rho_A}$ should match with the other two measures upto exponentially small terms. This result doesn't however constrain the Renyi entropies for a general Renyi index. As we will discuss below, it turns out that the volume law coefficient corresponding to Renyi entropies is same for all three measures. At the same time, as discussed   in detail in Sec.\ref{sec:numerics}, for finite sized systems, $S^A_n(\overline{\tr \rho^n_A}) $ is always a better measure of $S_{n,\textrm{avg}}$ compared to $S^A_n(\overline{ \rho_A}) $ due to the aforementioned result that their difference is exponentially small in the volume (see Fig.\ref{fig:s2_MBB_compare_avg}).

To begin with, let us briefly consider $S^A_n(\overline{ \rho_A}) =  \frac{1}{1-n} \log \left( \tr \left((\overline{\rho_A}) ^n\right)\right)$. 

\begin{equation}\label{eq:avgrhosn}
S^A_n(\overline{ \rho_A}) = \frac{1}{1-n}\log \left[\frac{   \sum_{E_A} e^{ S^M_A(E_A)+ nS^M_{\overline{A}}(E-E_A) }} {\left[\sum_{E_A} e^{S_A^M(E_A)+S_{\overline{A}}^M (E-E_A)} \right]^n} \right],
\end{equation}
where $S^M_A(E_A)$  denotes the logarithm of the density of states of $H_A$ at energy $E_A$. Similarly,  $S_{\overline{A}}^M (E-E_A)$ denotes the logarithm of the density of states of $H_{\overline{A}}$ at energy $E-E_A$.
Below, we will show that this expression matches that for $\overline{S^A_n}$ at the leading order in the thermodynamic limit when $V_A/V$ is held fixed.

For brevity, from now on we will drop the superscript `$A$' on the Renyi entropies $S^A_n$ for the rest of paper. To analyze $\overline{S_n}$, our main focus, let us first consider the second Renyi entropy $\overline{S_2}$. One finds (see Appendix \ref{sec:S2calculation}):

\begin{equation}\label{eq:s2}
\overline{S_2}= - \log \left[\frac{   \sum_{E_A} e^{ S^M_A(E_A)+ 2S^M_{\overline{A}}(E-E_A) }+ e^{ 2S^M_A(E_A)+S^M_{\overline{A}}(E-E_A) } }{   \left[\sum_{E_A}   e^{ S^M_A(E_A)+ S^M_{\overline{A}}(E-E_A) } \right]^2} \right].
\end{equation}
Unlike $S^A_n(\overline{ \rho_A})$, this expression is  manifestly symmetric between $A$ and $\overline{A}$. Most importantly, $\overline{S_2}$  is a \textit{universal} function of the microcanonical entropy (= logarithm of density of states) for the system. Furthermore, when $V_A/V<1/2$ is held fixed, in the thermodynamic limit (i.e. $V\to \infty$),  $\overline{S_2}$ can be simplified as 
  
  \begin{equation}\label{s2_first}
  \overline{S_2}= - \log \left[\frac{   \sum_{E_A}  e^{ S^M_A(E_A)+2S^M_{\overline{A}}(E-E_A) } }{   \left[\sum_{E_A}   e^{ S^M_A(E_A)+ S^M_{\overline{A}}(E-E_A) } \right]^2} \right].
  \end{equation}
  Let's consider the limit $V_A/V\to 0$. Taylor expanding $S^M_{\overline{A}}(E-E_A)$ as  $S^M_{\overline{A}}(E-E_A)=S^M_{\overline{A}}(E)-\beta E_A$, one finds
\begin{equation}\label{s2_small_f}
\overline{S_2}= - \log \left[\frac{ \tr e^{-2\beta H_A}  }{\left(\tr e^{-\beta H_A}\right)^2} \right]= - \log \left[\frac{ Z_A(2\beta)}{Z^2_A(\beta) }   \right]
\end{equation}
$=2\beta\left[F_A(2\beta) -F_A(\beta) \right].$ where $F_A(\beta)$ is the free energy of $H_A$ at temperature $\beta^{-1}$. This is exactly what one expects when the reduced density matrix is canonically thermal i.e. $\rho_A = e^{-\beta H_A}/\tr{e^{-\beta H_A}} $. Evidently, this result is true only when  $V_A/V\to 0$ and does not hold true for general values of $V_A/V$ and we will explore this and related aspects in much detail below.

Following the same procedure as above, one can also derive the universal formula for the Renyi entropy at an arbitrary Renyi index $n$. For example, the explicit expression for the third Renyi entropy is (Appendix \ref{sec:Sncalculation}):
\begin{widetext}
\begin{equation}\label{eq:s3}
\overline{S_3}=-\frac{1}{2}  \log \left[  \frac{   \sum_{E_A} e^{ S^M_A(E_A)+
		3 S^M_{\overline{A}}(E-E_A) }  + 3e^{ 2S^M_A(E_A)+
		2S^M_{\overline{A}}(E-E_A) }  +e^{ S^M_A(E_A)+
		S^M_{\overline{A}}(E-E_A) }  +e^{ 3S^M_A(E_A)+
		S^M_{\overline{A}}(E-E_A) }   }{ \left[\sum_{E_A}  e^{ S^M_A(E_A)+ S^M_{\overline{A}}(E-E_A) } \right]^{3}  }      \right]
\end{equation}
\end{widetext}

The explicit expression of $n$'th Renyi entropy can be expressed as a logarithm of the sum of $n!$ terms. In the thermodynamic limit, however, only one of these terms is dominant, and the expression becomes (for $V_A/V <1/2$):

\begin{equation}\label{eq:sn}
\overline{S_n}= \frac{1}{1-n}\log \left[\frac{   \sum_{E_A} e^{ S^M_A(E_A)+ nS^M_{\overline{A}}(E-E_A) }}{   \left[\sum_{E_A}   e^{ S^M_A(E_A)+ S^M_{\overline{A}}(E-E_A) } \right]^n} \right].
\end{equation}

Note that this is identical to the Renyi entropy  $S^A_n(\overline{ \rho_A})$, Eq.\ref{eq:avgrhosn}. See Appendix \ref{sec:Sncalculation} for  details of the calculation.
%, which is a consequence of equivalence of reduced density matrix between a mixed state and a pure state for $V_A/V<1/2 $ in thermodynamic limit.

\subsection{Curvature of  Renyi Entropies and the Failure of Page Curve} \label{sec:curvature}

Let us evaluate Eq.\ref{eq:sn}, in thermodynamic limit $V\to \infty$ with $f=V_A/V (<1/2)$ held fixed. The thermodynamic limit allows one to use the saddle point approximation technique. The numerator can be written as,
\begin{equation}
\begin{split}
\sum_{E_A} e^{ S^M_A(E_A)+nS^M_{\overline{A}}(E-E_A) }&=\sum_{u_A} e^{ V_As(u_A)+nV_{\overline{A}}s(u_{\overline{A}}) }\\
\end{split}
\end{equation}
where $u_A$ denotes the energy density in $A$ while $u_{\overline{A}}$ denotes the energy density in $\overline{A}$ consistent with energy conservation, and $s(u)$ is the entropy density at energy density $u$. Thus, 
\begin{equation}\label{eq:e_conservation}
u_{\overline{A}}=\frac{u}{1-f}-\frac{f}{1-f}u_A.
\end{equation}
where $u = E/V$ is the energy density corresponding to the eigenstate under consideration. At the saddle point, the sum over $u_A$ is dominated by the solution to the equation:
\begin{equation}\label{eq:saddle}
\boxed{ \frac{\partial s(u)}{\partial u} \biggr\rvert_{u=u^*_A} =n\frac{\partial s(u)}{\partial u } \biggr\rvert_{u=u^*_{\overline{A}}} }
\end{equation}
and therefore the numerator equals  $ e^{ V \left[   fs(u^*_A)+n(1-f)s(u^*_{\overline{A}})  \right]}$ in thermodynamic limit.

On the other hand, the denominator is 

\begin{equation}
\begin{split}
\sum_{E_A} e^{ S^M_A(E_A)+S^M_{\overline{A}}(E-E_A) }&=\sum_{u_A} e^{ V_As(u_A)+V_{\overline{A}}s(u_{\overline{A}}) }\\
&=e^{Vs(u)},
\end{split}
\end{equation}
where we have used the fact that the saddle point for the denominator is $u_A^*=u_{\overline{A}}^*=u$, i.e., it is unchanged from the energy density of the eigenstate under consideration.

Combining the above results, $\overline{S_n}$ is therefore given by:

\begin{equation}\label{eq:sn_saddle}
\boxed{\overline{S_n}=\frac{V}{1-n}\left[ fs(u_A^*)  +n(1-f)s(u^*_{\overline{A}}) -ns(u)  \right]}
\end{equation}
where $u_A^*$ and $u_{\overline{A}}^*$ are obtained by solving  the saddle point condition Eq.\ref{eq:saddle}. 

This is the central result of our paper. Several observations can be made immediately:

\vspace{0.3cm}

1. When $n=1$, $u^*_A = u$ i.e. the von Neumann entanglement entropy $S_1$ depends only on the density of states at the energy density corresponding to the eigenstate \textit{for all values of $f = V_A/V$}. Furthermore, the volume law coefficient of $S_1$ is strictly linear with $V_A$, i.e., $S_1 = s(u) V_A$ for $f < 1/2$. We will call such linear dependence `Page Curve' \cite{page1993average,lubkin1978}, as is conventional. As discussed in the Introduction, this result was also argued for in Ref.\cite{garrison2015does} and Ref.\cite{dymarsky2016subsystem}.

\vspace{0.3cm}

2. When $n \neq 1$, the  Renyi entropy density $S_n/V_A$ as $V \to \infty$ for fixed $ f= V_A/V$ depends on $f$, and thus the Renyi entropies have a non-trivial curvature dependence when plotted as a function of $V_A/V$. Perhaps most interestingly, as shown in Appendix \ref{sec:saddle}, the curvature $\dfrac{d^2S_n}{d f^2}$ depends only on the sign of $n-1$:

\begin{equation}
\begin{split}
&\overline{S_n(f)} \quad \text{is convex for } \quad n>1\\
&\overline{S_n(f)}  \quad \text{is concave for } \quad n<1.
\end{split}
\end{equation}

\vspace{0.3cm}

3.  The saddle point equation (Eq.\ref{eq:saddle}) implies that for a fixed Renyi index $n$, the energy density $u^*_A$ that determines the volume law coefficient of $\overline{S_n}$ depends on $f$. Therefore, different values of $f$ encode thermodynamical information at different temperatures. Recall that in contrast, as $f \rightarrow 0$, the $n$'th Renyi entropy depends only on the free energy densities at temperature $\beta^{-1}$ and $(n \beta)^{-1}$.

\vspace{0.3cm}

We recall that the Renyi entanglement entropies $S_n$ corresponding to a \textit{ typical} state in the Hilbert space \cite{page1993average,lubkin1978,LLOYD1988,sen1996,sanchez1995}  equals $\log(\mathcal{H}_A)$ where $\mathcal{H}_A$ is the size of the Hilbert space in region $A$ (assuming $\mathcal{H}_A < \mathcal{H}_{\overline{A}} $). For a system with a local Hilbert space dimension $\mathcal{H}_{\textrm{local}}$, this translates as a volume law for Renyi entropies i.e. $S^n_A = V_A \log(\mathcal{H}_{\textrm{local}})$ as long as $f < 1/2$ (e.g. in a spin-1/2 system, $S^n_A = V_A \log(2)$). This result matches the entropy corresponding to a thermal ensemble at infinite temperature. Based on this, one might have  expected that for an eigenstate of a physical Hamiltonian at temperature $\beta^{-1}$,  the Renyi entropies are perhaps given by their canonical counterparts i.e. $S_n = V_A n\beta (\mathfrak{f}(n \beta) -  \mathfrak{f}(\beta))/(n-1)$ for all $f <1/2$,  a finite temperature version of Page Curve ($\mathfrak{f}(\beta)$ is the free energy density). \textit{Our result indicates that this is not the case, and Renyi entropies for  $n\neq 1$ do not follow such a Page Curve}.

\vspace{0.4cm}

%\centerline{\textbf{{ An example: Renyi Entropy for System with  \newline Gaussian Density of States}}}~\\

\centerline{\textbf{An Example:}}
\centerline{\textbf{ Renyi Entropy for System with  Gaussian Density of States}}
 
 \vspace{0.2cm}
 
Let's study an example where one can solve the saddle point  Eq.\ref{eq:saddle}, and solve for the Renyi entropies explicitly. Consider a system with volume $V$ where  the density of states $g(E)$ is a Gaussian as a function of the energy $E$:
\begin{equation}
g(E)= e^{V\log 2-\frac{E^2}{2V}}, 
\end{equation}
Thus, the microcanonical entropy density is given by 
\begin{equation}\label{eq:s(u)_gaussian}
s(u)=\log 2 -\frac{1}{2}u^2
\end{equation}
where $u\equiv E/V$ denotes the energy density. This expression also implies that the temperature $\beta(u) = - u$. As a practical application, all systems whose energy-entropy relation $s(u)$ is symmetric under $u \rightarrow -u$, a Gaussian density of states will be a good approximation to the function $s(u)$ close to the infinite temperature. Therefore, the results derived can be thought of as a leading correction to the Renyi entropy in a high temperature series expansion for such systems.

Directly evaluating the expression in Eq.\ref{eq:s2}, one finds the following expression for $S_2$ (see Appendix \ref{sec:Gaussian_derivation}):
\begin{equation}\label{eq:s2_gaussian}
\overline{S_2}=-\log\left[\frac{1}{\sqrt{1-f^2}}e^{-V\gamma(f,u)} +\frac{1}{\sqrt{1-(1-f)^2}}e^{-V\gamma(1-f,u)} \right],
\end{equation}

where

\begin{equation}
\gamma (f,u)=f\log 2 -\frac{f}{1+f}u^2
\end{equation}
When $0<f<\frac{1}{2}$ ( $\frac{1}{2} < f < 1$), the first (second) term dominates in the thermodynamic limit.Thus, for  $0<f<\frac{1}{2}$,
\begin{equation}
\overline{S_2}=fV\left(\log2-\frac{u^2}{1+f}\right)=fV\left(\log2-\frac{\beta^2}{1+f}\right).
\end{equation}

\begin{figure}
	\centering
	\begin{subfigure}[b]{0.35\textwidth}
\captionsetup{justification=centering,singlelinecheck=false}
		\includegraphics[width=\textwidth]{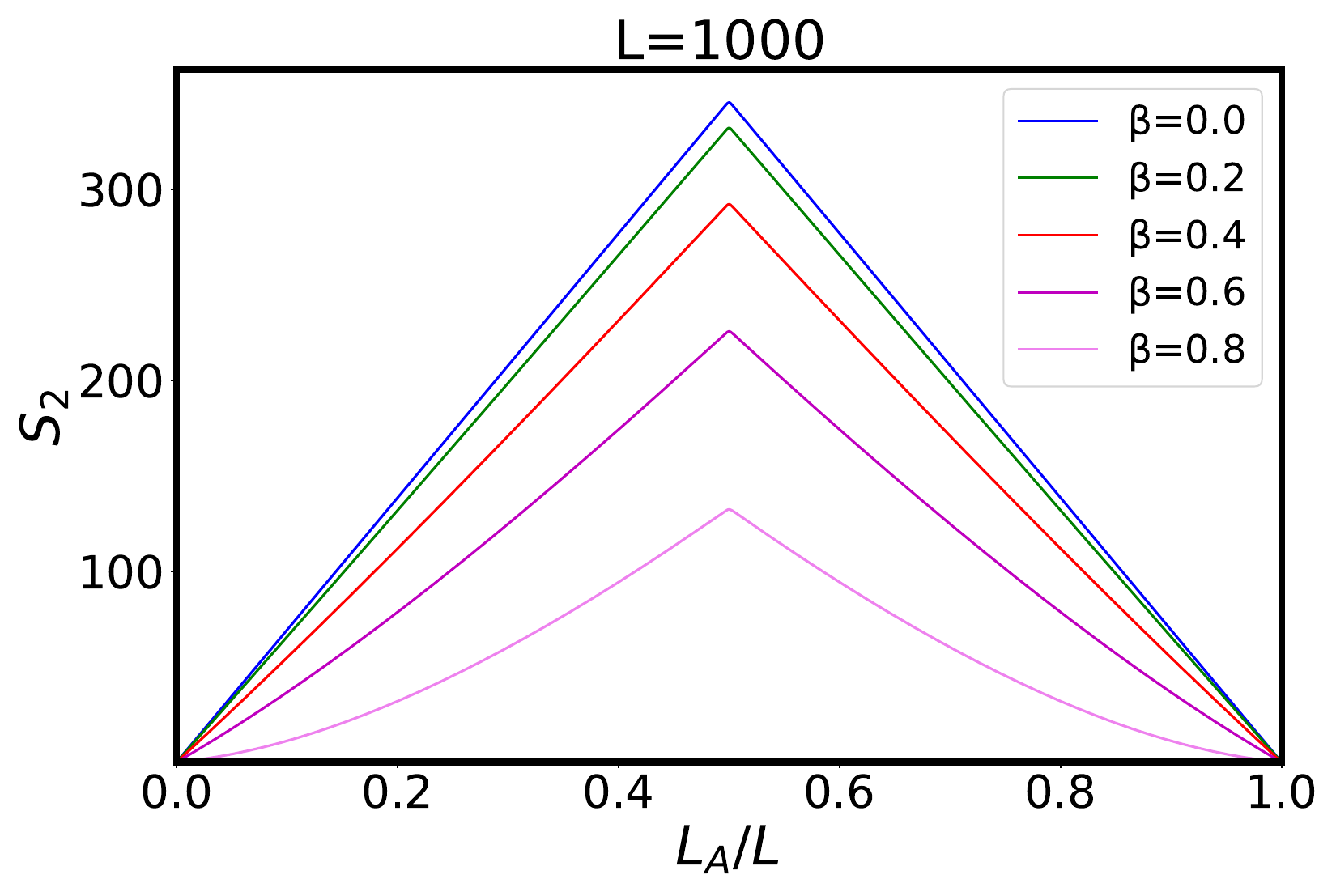}
		\caption{$\overline{S_2}$}
		\label{fig_s2}
	\end{subfigure}
	\begin{subfigure}[b]{0.35\textwidth}
		\captionsetup{justification=centering,singlelinecheck=false}
		\includegraphics[width=\textwidth]{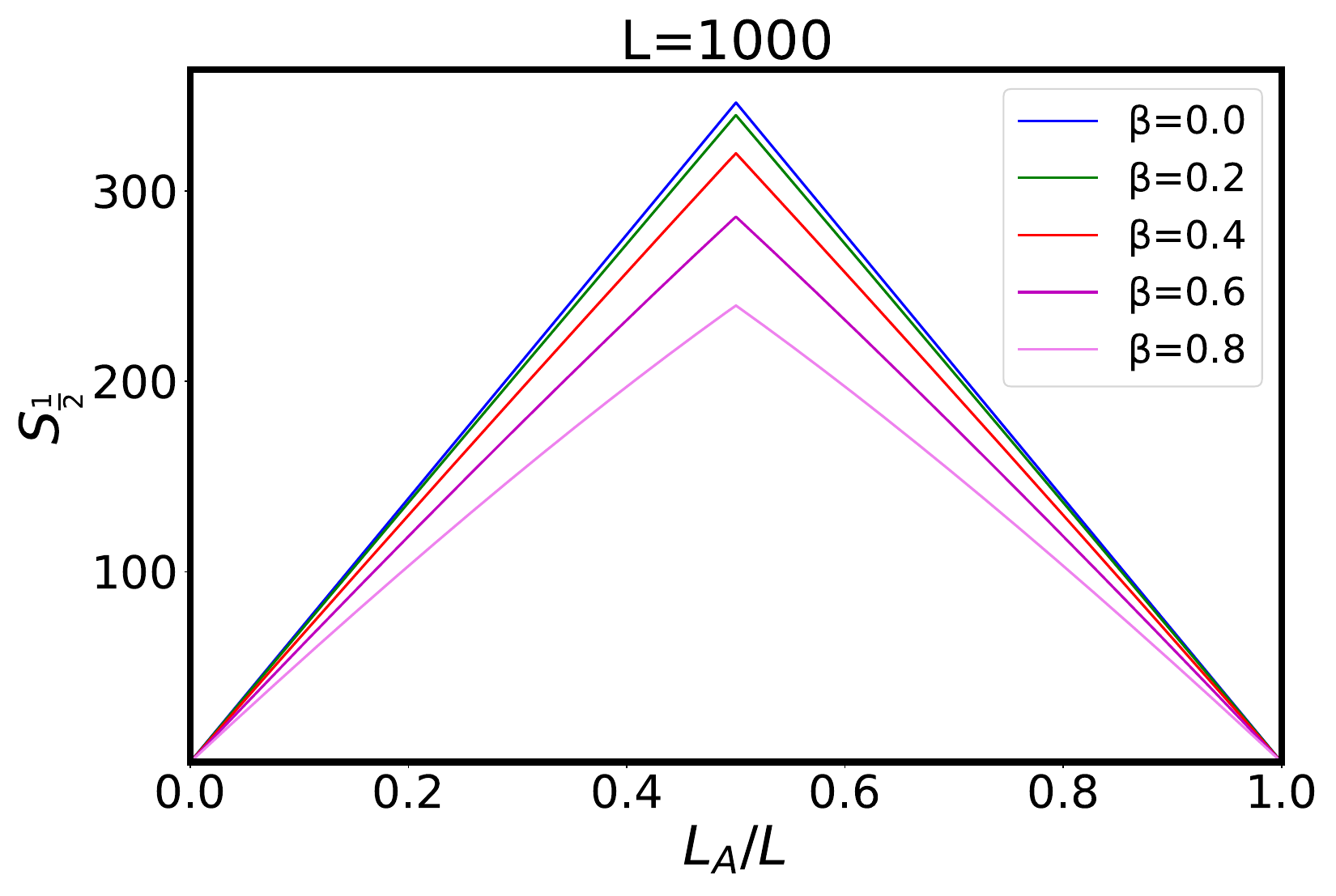}
		\caption{$\overline{S_{1/2}}$}
		\label{fig_s1_2}
	\end{subfigure}
	\caption{The Renyi entropies $S_2$ (top) and $S_{1/2}$ (bottom) for a system with Gaussian density of states  (Eq.\ref{eq:sn_gaussian}). } 
\end{figure}

Similarly, one can obtain Renyi entropy for arbitrary Renyi index $n$ for  $0<f<\frac{1}{2}$ in the thermodynamic limit:

\begin{equation}\label{eq:sn_gaussian}
\overline{S_n}=fV\left[\log2-\frac{n\beta^2}{2(1+(n-1)f)}\right].
\end{equation}
This expression illustrates several of the general properties discussed in the previous subsection. First we notice that $\overline{S_n}$ is linear for arbitrary $\beta $ only when $n=1$, and therefore the von Neumann entropy follows the finite temperature Page curve. For $n\neq 1$, $\overline{S_n}$ is linear in $f$ only at the infinite temperature, and the non-linear dependence on $f$ becomes non-negligible as one moves away  from the infinite temperature. Furthermore, the Renyi entropies are convex functions of $V_A$  for $n>1$ while they are concave for $n<1$. As a demonstration, we plot Eq.\ref{eq:sn_gaussian} for different $\beta$ with $n>1$ and $n<1$ respectively in Fig.\ref{fig_s2} and Fig.\ref{fig_s1_2}, where we clearly observe the concave and convex shape for Renyi entropies. ~\\

\subsection{Comparison with `Pure Thermal' State} \label{sec:compare_thermal}

 Recently, Ref. \cite{fujita2017universality} also studied the entanglement entropies of chaotic systems using an approach which is similar in spirit to ours, but for a different class of states. They considered a ``canonical thermal pure quantum (CTPQ) '' state:
\begin{equation}
\ket{\psi}=\frac{1}{\tr e^{-\beta H}}\sum_j z_j e^{-\beta H/2} 	\ket{j}   
\end{equation}
where $\{  \ket{j}  \}$ form a complete orthonormal bases in the Hilbert space, and the coefficient $z_j$ is a random complex number $z_j\equiv \left( x_j+iy_j\right)/\sqrt{2} $ with $x_j$ and $y_j$ is i.i.d based on  a Gaussian with zero mean and unit variance. They calculated the Renyi entropy of the CTPQ states and used the functional form thus obtained as a fitting function for Renyi entropies of chaotic eigenstates obtained via exact diagonalization. For reference, we write down the expression of second Renyi entropy obtained in their paper:
\begin{equation}\label{eq:fujita}
\overline{S_{2,CTPQ}} =  -\log \frac{\tr_A\left(\tr_B e^{-\beta H}\right)^2  +\tr_B\left(\tr_A e^{-\beta H}\right)^2    }{\left(\tr e^{-\beta H}\right)^2}. 
\end{equation}
Note the resemblance with our result Eq.\ref{eq:s2}. Despite the apparent similarity, the  functional dependence of Renyi entropy obtained from Eq.\ref{eq:fujita} is actually quite  different than our result, Eq.\ref{eq:sn_saddle}. In particular, for fixed $ V_A/V$ (<1/2), as $V\to \infty $, one may verify that the volume law coefficient of the Renyi entropy  $\overline{S_{n,CTPQ}}$ corresponding to the CTPQ state actually matches that of a thermal state: $\overline{S_{n,CTPQ}}= \frac{n}{n-1} V_A \beta \left(  \mathfrak{f}(n \beta) - \mathfrak{f}(\beta) \right)$, and therefore follows the Page Curve. This is in contrast to the MBB/EB states, which as discussed above, have a distinct curvature dependence.  One may also verify that the reduced density matrix in region $A$ of a CTPQ state: 
\begin{equation}
\rho_A \sim \frac{e^{-\beta H}}{\tr e^{-\beta H}}
\end{equation}
\textit{for any $V_A/V$} in thermodynamic limit which implies that the energy variance $\sim V_A$ for all $V_A/V$ and does not respect the fact that for an eigenstate, the energy variance should be symmetric around $V_A/V  = 1/2$ (similar to Renyi entropies), and should vanish  when $V_A/V \rightarrow 1$.

\section{Comparison of Analytical Predictions with Exact Diagonalization} \label{sec:numerics}

\begin{figure}
	\centering
	\includegraphics[width=0.5\textwidth]{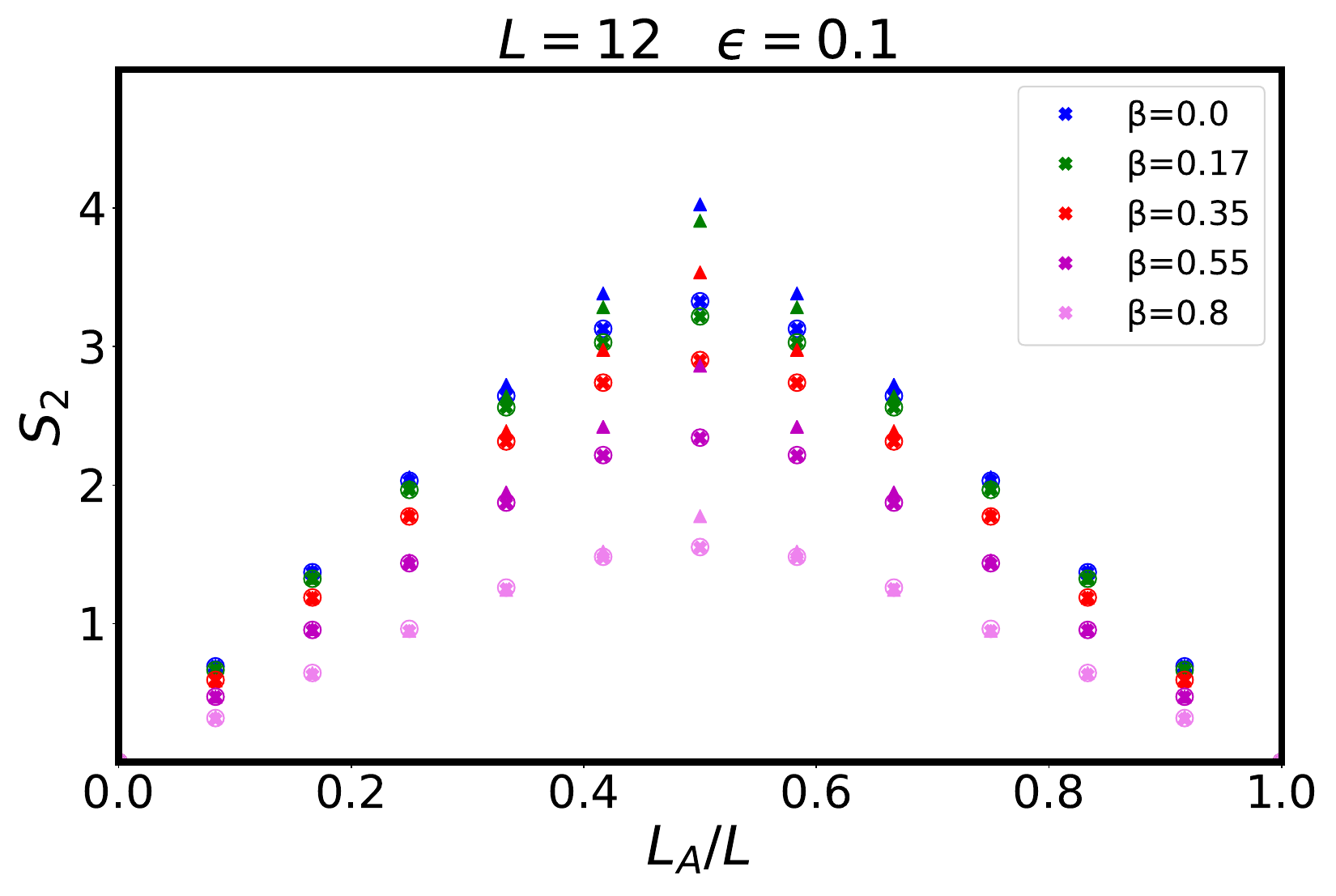}
	\caption{Comparison of the three different ways to average over the random ensembles discussed in the text to obtain the second Renyi entropy. Triangles: $S^A_2(\overline{ \rho_A})$. Crosses: $S^A_2(\overline{\tr \rho^2_A})$. Open circles: $S^A_{2,\textrm{avg}} = - \overline{\log \left( \tr \left(\rho^2_A \right)  \right)}$. Note that $S^A_2(\overline{\tr \rho^2_A})$ and $S^A_{2,\textrm{avg}}$ are essentially identical, as they should be (see Appendix \ref{sec:equiv} ).  The Hamiltonian is $H=-\sum_{i=1}^N Z_i  +\epsilon H_1$, where $H_1$ is a real hermitian random matrix. } 
	\label{fig:s2_MBB_compare_avg}
\end{figure}

In this section, we will compare our analytical predictions with numerical simulations on quantum spin-chain Hamiltonians. Recall that our analytical results are for $\overline{S_n}  \stackrel{def}{=} \frac{1}{1-n} \log\left( \overline{\tr \rho^n_A}\right)$, which is essentially identical to the more physical quantity, $S_{n,\textrm{avg}} =  \frac{1}{1-n} \overline{\log \left( \tr \left(\rho^n_A \right)  \right)}$, as discussed at the beginning of Sec.\ref{sec:renyi_dos} and in Appendix \ref{sec:equiv}). See Fig.\ref{fig:s2_MBB_compare_avg} for a demonstration. Due to this, we will continue to use the symbol $\overline{S_n} $  for Renyi entropies obtained from  numerical simulations even though we are really calculating $S_{n,\textrm{avg}} $. In contrast, the quantity $S_n(\overline{ \rho_A}) =  \frac{1}{1-n} \log \left( \tr \left((\overline{\rho_A}) ^n\right)\right)$ which incidentally equals the asymptotic expression for $\overline{S_n}$ in the thermodynamic limit (see Eqs. \ref{eq:avgrhosn} and \ref{eq:sn}), does not agree as well with   $S_{n,\textrm{avg}}$ (Fig.\ref{fig:s2_MBB_compare_avg}).

We will   compare the ED results with the analytical results for MBB, EB and CTPQ states. Our approach will be different than the one in Ref.\cite{fujita2017universality} where the analytical results for the CTPQ state were used only as a guide to fit the results of ED.

\subsection{Non-integrable Spin-1/2 Chain Close to Integrable Regime}

In this subsection we numerically study Renyi entropies for eigenstates of a non-integrable Hamiltonian close to the classical limit, namely the Hamiltonians of the form in Eq.\ref{eq:MBB_H}:
\begin{equation}
H=H_0+\epsilon H_1,  \label{eq:perturbH}
\end{equation}
where $H_0$ denotes the classical, integrable local Hamiltonian and $\epsilon\ll O(1)$ is an integrability-breaking parameter. ~\\

\vspace{0.3cm}

\centerline{\textbf{\underline{Spin-$1/2$ Chain with Local Perturbation}}}~\\

\vspace{0.3cm}

Consider 
\begin{equation}
\begin{split}
&H_0=\sum_{i}^L  -Z_iZ_{i+1} -Z_i  \\
&H_1=\sum_i^L X_i,
\end{split}  \label{eq:Hberry}
\end{equation}

\begin{figure}
	\centering
	\begin{subfigure}[b]{0.23\textwidth}
		\captionsetup{justification=centering,singlelinecheck=false}
		\includegraphics[width=\textwidth]{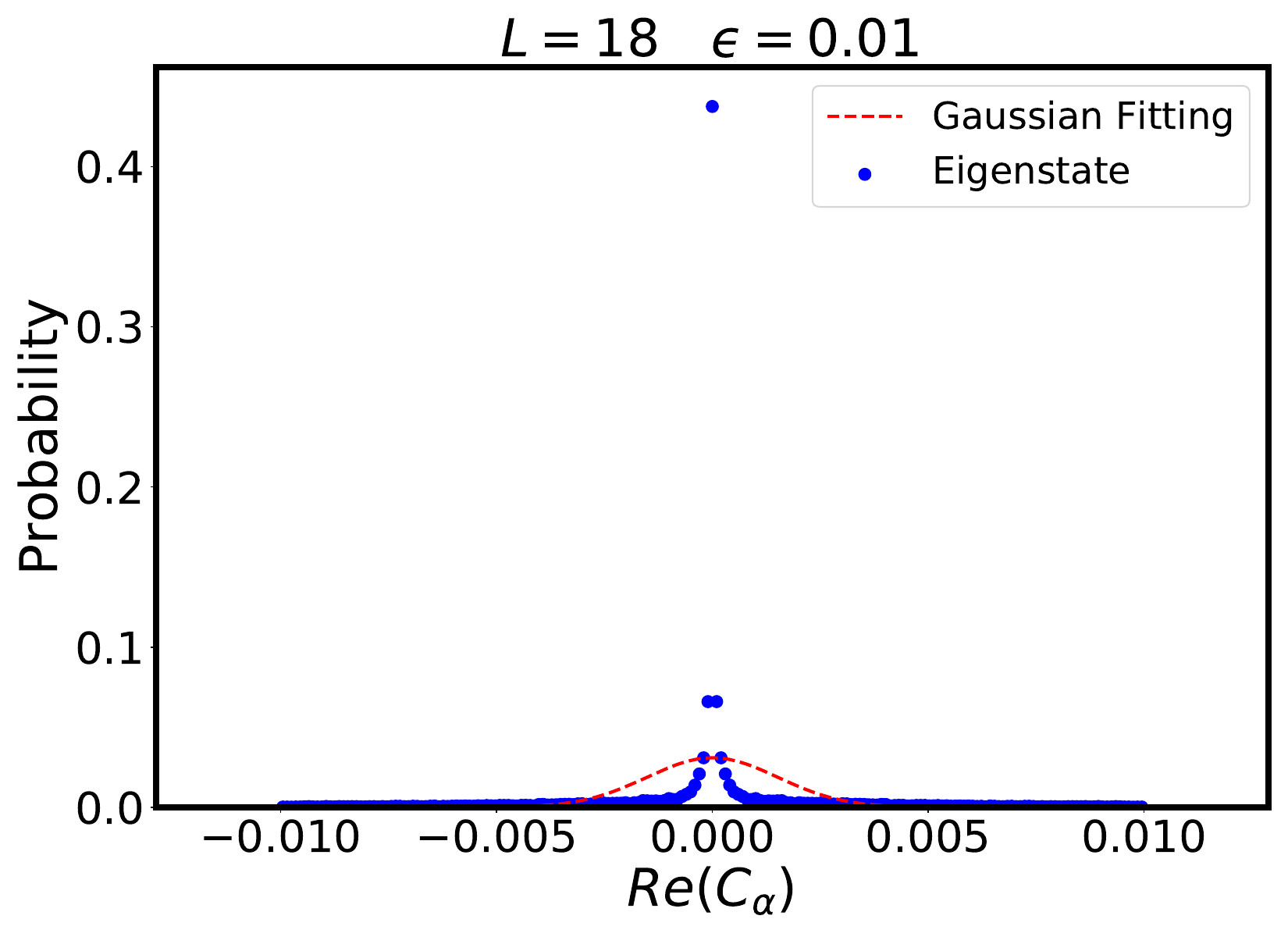}
		\caption{$\epsilon=0.010$}
	\end{subfigure}
	\begin{subfigure}[b]{0.23\textwidth}
		\captionsetup{justification=centering,singlelinecheck=false}
		\includegraphics[width=\textwidth]{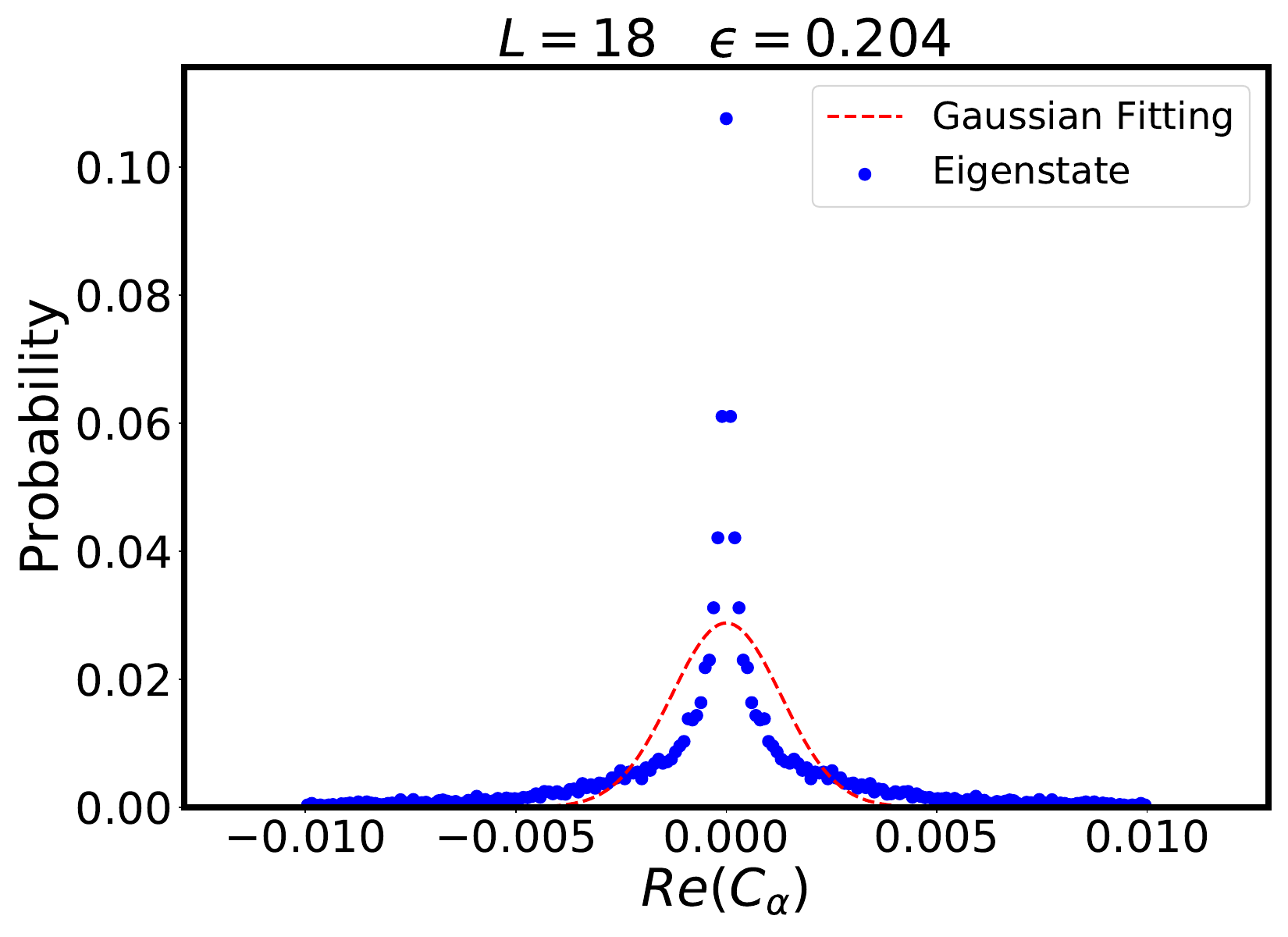}
		\caption{$\epsilon=0.204$}
	\end{subfigure}

	\begin{subfigure}[b]{0.23\textwidth}
		\captionsetup{justification=centering,singlelinecheck=false}
		\includegraphics[width=\textwidth]{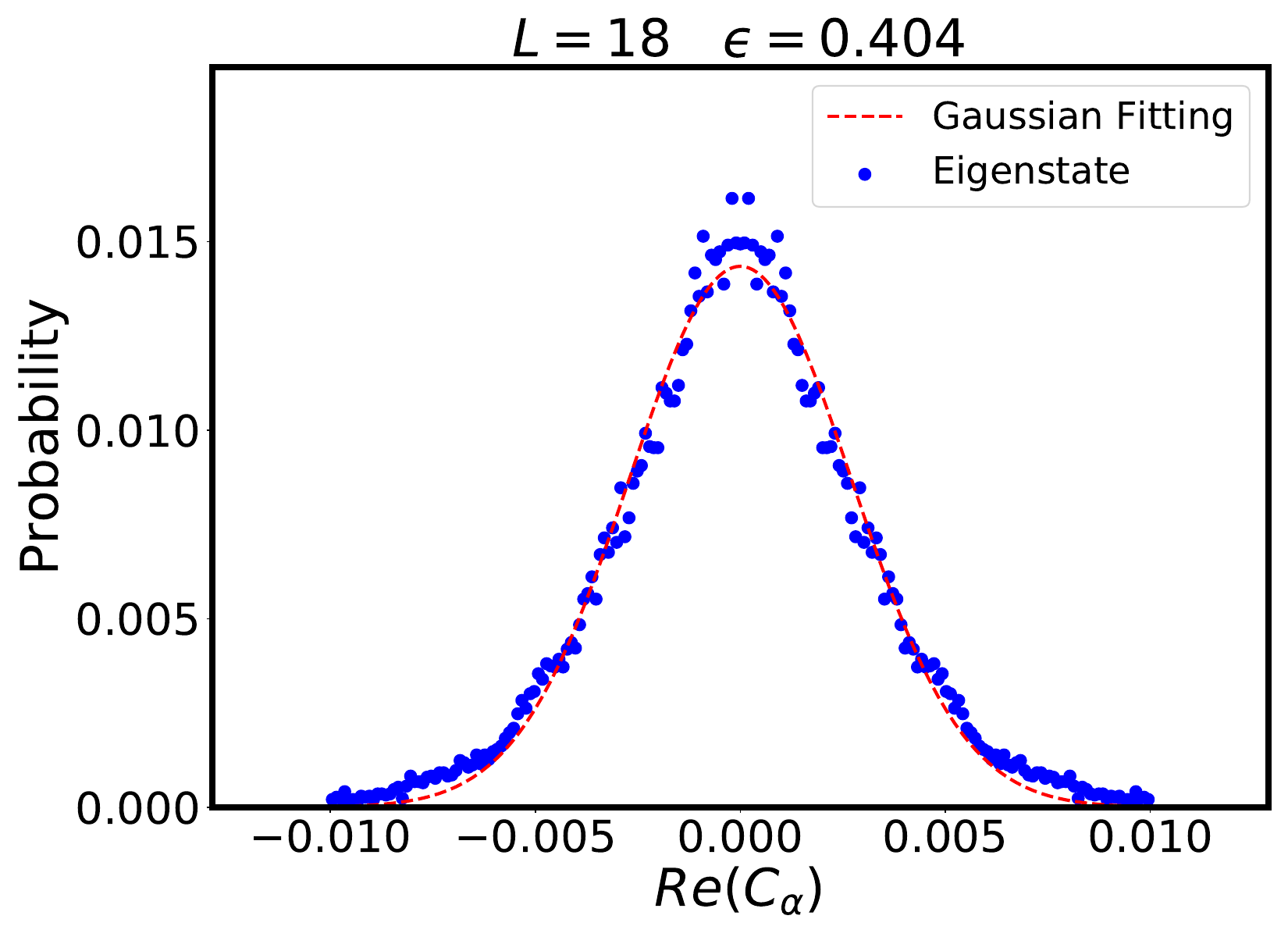}
		\caption{$\epsilon=0.404$}
	\end{subfigure}
	\begin{subfigure}[b]{0.23\textwidth}
		\captionsetup{justification=centering,singlelinecheck=false}
		\includegraphics[width=\textwidth]{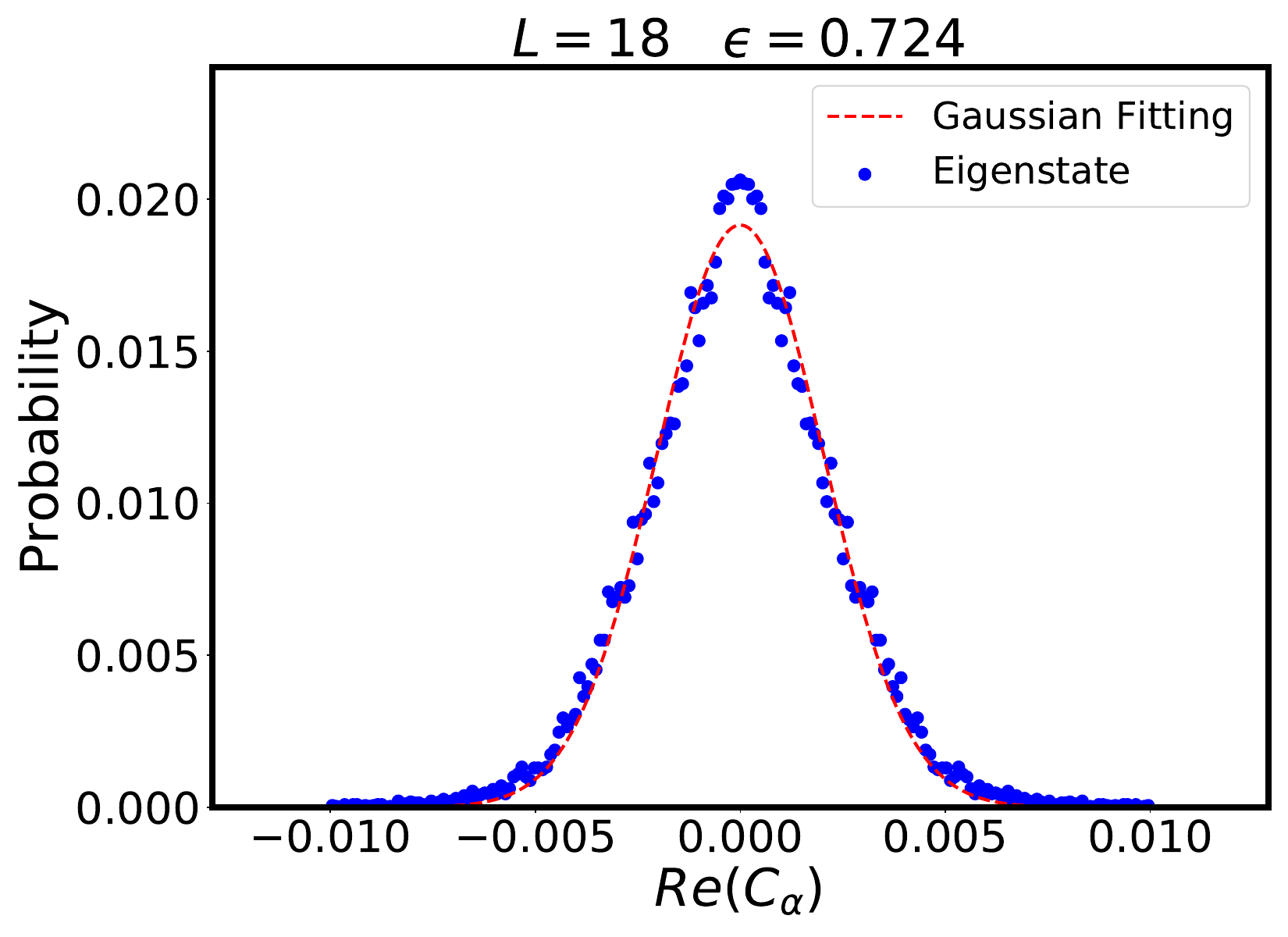}
		\caption{$\epsilon=0.724$}
	\end{subfigure}
	\caption{Probability amplitudes for $C_\alpha$ (Eq. \ref{eq:random_state}) for various values of the perturbation $\epsilon$ corresponding to the Hamiltonian in Eq.\ref{eq:perturbH} when $H_0$ and $H_1$ are given by Eq.\ref{eq:Hberry}. We choose  eigenstates at energy $E=0$ and the width of energy window is $\Delta =1$. The Gaussian distribution is obtained by least square fitting.   }
	\label{fig:amp_MBB}
\end{figure}

\begin{figure}
	\centering
	\includegraphics[width=0.5\textwidth]{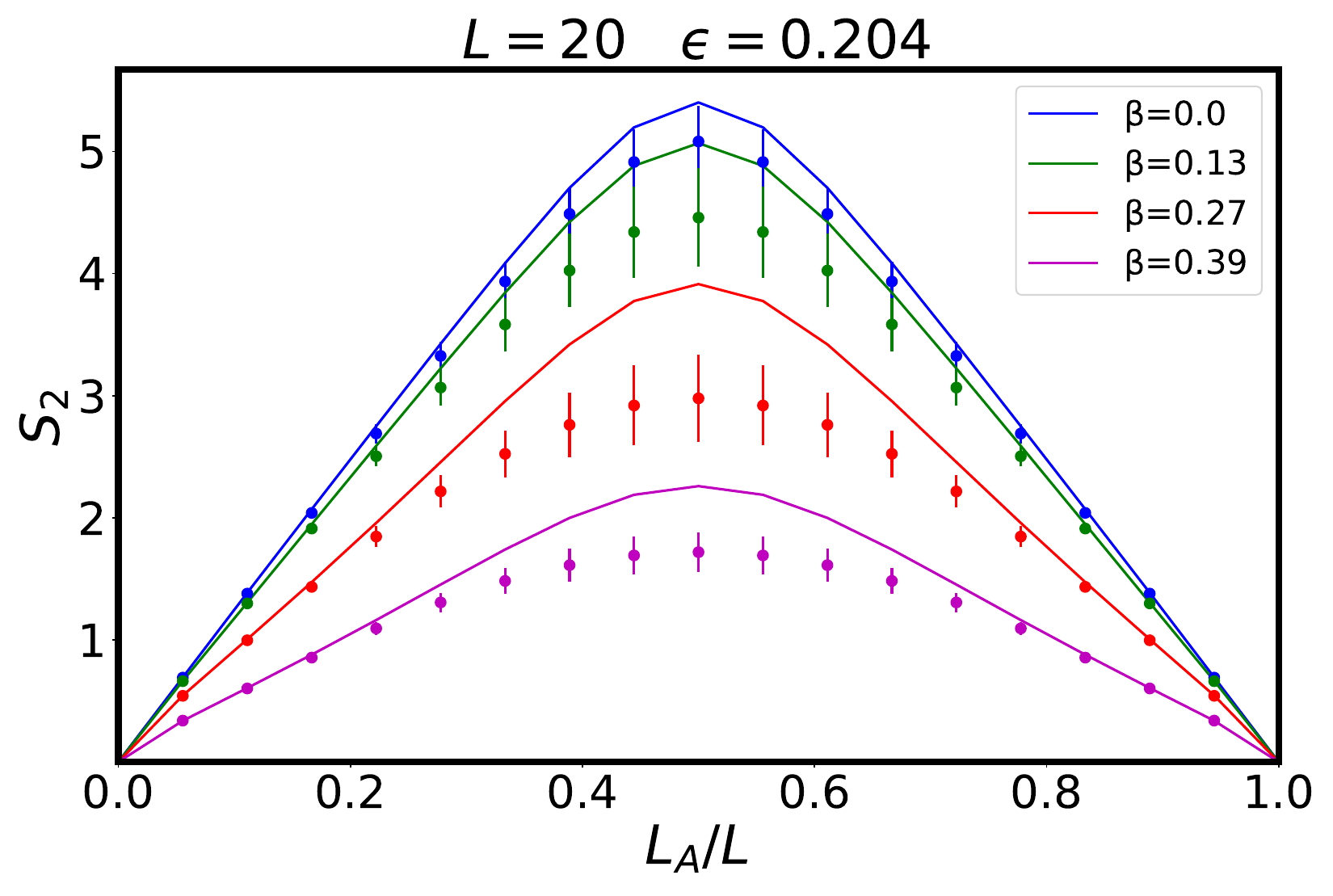}
	\caption{Comparison of the second Renyi entropy $S_2$ obtained from the many-body Berry conjecture, with those obtained from the exact diagonalization for the Hamiltonian $H=H_0+\epsilon H_1$ where $H_0$ and $H_1$ are given by Eq.\ref{eq:Hberry}. The solid dots correspond to  $S_2$ of eigenstates averaged over an energy window of width $\Delta E=2$, and the vertical bars denote the standard deviation in $S_2$ in this energy window. Solid lines correspond to $S_2$ for the MBB state using Eq.\ref{eq:s2}.} 
	\label{fig:s2_MBB_spin_chain}
\end{figure}

We first study the histogram of amplitudes $C_\alpha$ introduced in Eq.\ref{eq:random_state} for various values of $\epsilon$ to check the validity of Many-body Berry (MBB) conjecture. In Fig.\ref{fig:amp_MBB}, we observe that amplitudes approach a Gaussian probability distribution with increasing $\epsilon$. Analytical and numerical estimates suggest that one requires $\epsilon \gtrsim 1/L^{\beta}$ where $\beta$ is some positive number to access the chaotic regime  \cite{rigol2010quantum, flambaum1997criteria, georgeot2000quantum,santos2012chaos,mukerjee2014,srednicki_kitp}.  Evidently (Fig.\ref{fig:amp_MBB}), due to system size limitations, one requires $\epsilon \approx 0.4$ to really see the onset of chaos in our simulations. Therefore, one doesn't expect that the eigenstates of $H$  in the chaotic regime  can  be obtained solely by randomly superposing eigenstates of $H_0$, and we are unable to verify the MBB conjecture for this system. Fig.\ref{fig:s2_MBB_spin_chain} compares the  Renyi entropy of the eigenstates of $H$ with those predicted by MBB conjecture when $\epsilon$ is smaller than $\approx 0.4$. Curiously, although we are not able to predict the full shape dependence of Renyi entropy using the MBB conjecture for the reasons just outlined, it still works rather well to predict the Renyi entropies for $V_A/V \ll 1$. This is perhaps not surprisingly, since physically, the cross-over value of $\epsilon$ required to obtain chaos at smaller length scales should be smaller than the one required for the whole system. ~\\

\vspace{0.3cm}

\centerline{\textbf{\underline{Spin-$1/2$ Chain with Random Non-local  Perturbation}}}

\vspace{0.2cm}

Our expectation is that the system size at which the cross-over  from integrability to chaos occurs is parametrically smaller when $H_1$ is non-local as compared to when it is local. In fact, a diagonal $N \times N$ matrix perturbed by a matrix chosen from a random Gaussian Orthogonal Ensemble (GOE) shows chaotic behavior when the strength of the perturbation  \cite{zirnbauer1983, mcgreevy} $\gtrsim 1/\sqrt{N}$. Translating this to the many-body Hamiltonians with  Hilbert space size $\mathcal{H}$, this indicates a cross-over scale of  $1/\sqrt{\mathcal{H}} = 2^{-L^d/2}$. 
 
Consider  $H = H_0 + \epsilon H_1$ where 

\begin{equation}\label{eq:nonlocalH}
H_0=-\sum_{i=1}^L Z_i
\end{equation}
and $H_1$ is chosen  randomly  from the GOE.  The variance corresponding to the probability distribution function of the matrix elements in $H_1$ is chosen such that the range of energy spectrum of $H_1$  is $L$. 

We again emphasize that despite the non-locality of $H$, the MBB states depend \textit{solely} on $H_0$, which \textit{is} local. Due to this, the MBB states continue to satisfy properties expected from a local Hamiltonian, such as the validity of cluster decomposition of correlations of local operators.

The advantage of working with the above $H_0$ is that one can calculate its density of states exactly, and therefore obtain analytical predictions for the Renyi entropies of the chaotic Hamiltonian $H$. In particular, the number of eigenstates of $H_0$ at energy $E$ are:
  
\begin{equation}
g=\frac{L!}{N_{\uparrow}!N_{\downarrow}!}%=\frac{L!}{\left(\frac{L-E}{2}\right)!\left(\frac{L+E}{2}\right)!  }. 
\end{equation} 

where $N_{\uparrow}=\frac{L-E}{2}$ and $N_{\downarrow}=\frac{L+E}{2}$. Thus the microcanonical entropy $S^M = \log(g)$  under Sterling's approximation is given by,

\begin{equation}
S^M \approx L\log(L)-\frac{L-E}{2} \log\frac{L-E}{2} -\frac{L+E}{2} \log\frac{L+E}{2}.
\end{equation}

In fact, at high temperatures, the entropy density is same as that of the Gaussian model,  Eq.\ref{eq:s(u)_gaussian}, $s=S^M/L=  \log2 -\frac{1}{2}\beta^2$.

\begin{figure}
	\centering
		\includegraphics[width=0.5\textwidth]{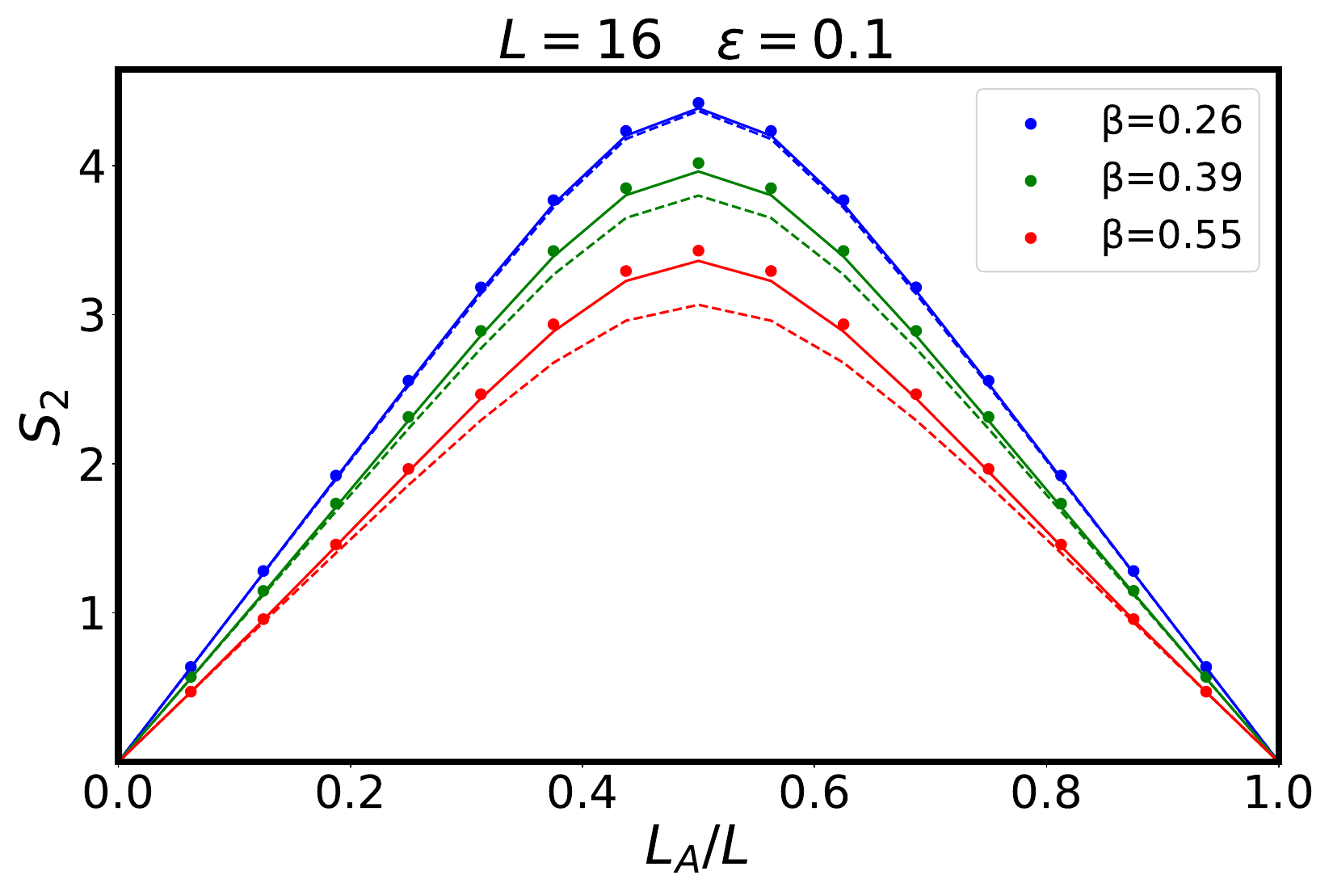}
	\caption{Comparison of the second Renyi entropy $S_2$ of 1D spin model with Hamiltonian $H = H_0 + \epsilon H_1$ where $H_0$ is given by Eq.\ref{eq:nonlocalH} and $H_1$ is a random matrix. The three plotted quantities correspond to Renyi entropy $S_2$ of eigenstates (solid dots), MBB states (Eq.\ref{eq:s2})(solid lines), and CTPQ states (Eq.\ref{eq:fujita}) (dashed lines).}
			\label{fig:s2_with_ctpq}
\end{figure}

Fig.\ref{fig:s2_with_ctpq} shows the comparison of the Renyi entropies of the eigenstates of $H$ at $\epsilon=0.1$ with the analytical predictions for an MBB state. We see that agreement is quite well for a wide range of temperatures.

In Fig.\ref{fig:s2_with_ctpq}, we also compare the results with the expression obtained from a  CTPQ state,  Fig.\ref{fig:s2_with_ctpq}. We see that they match well for small values of $f = L_A/L$. One the other hand, for $f = O(1)$, the Renyi entropy of a CTPQ state is smaller than the exact diagonalization results and the predictions from MBB state. This is consistent with the fact that in the thermodynamic limit, a CTPQ state predicts linear dependence of the second Renyi entropy as a function of $f$, while for an MBB state, the second Renyi entropy is a convex function of $f$ (Sec.\ref{sec:curvature}).

\begin{figure}
	\centering
	\includegraphics[width=0.5\textwidth]{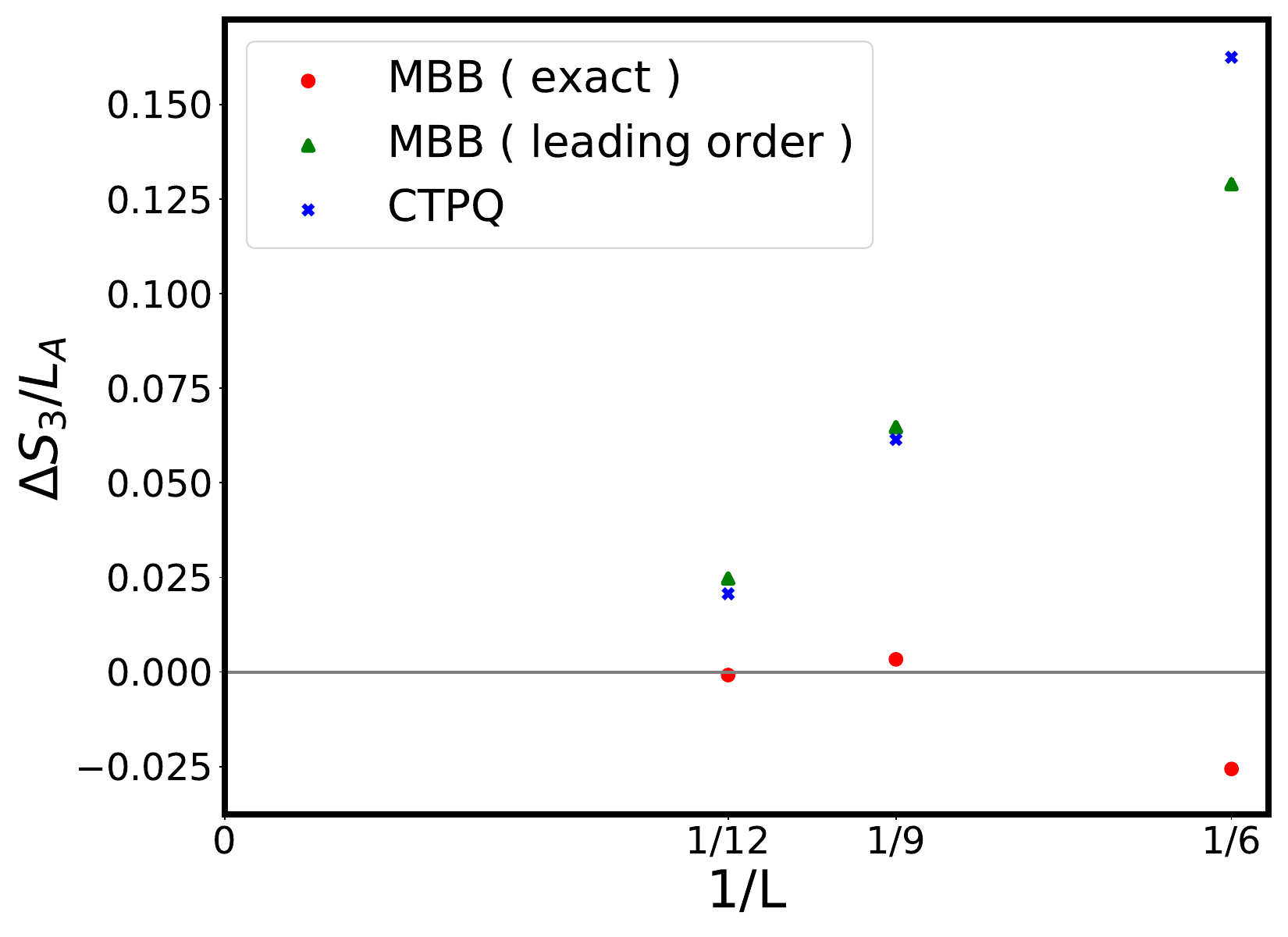}
	\caption{Finite size scaling of the difference between the third Renyi entropy $S_3$ obtained from anaytical expressions and exact diagoanlization results for the eigenstates of a 1D spin model with Hamiltonian $H = H_0 + \epsilon H_1$ where $H_0$ is given by Eq.\ref{eq:nonlocalH} and $H_1$ is a random matrix. Here $\epsilon =0.1$. A data point is obtained by averaging $\Delta S_3$ for eigenstates in the range $0<\beta<0.5 $. $L_A $ is chosen to be $L/3$.  } 
	\label{fig:finite_size_MBB_s3_av}
\end{figure}

\vspace{0.4cm}

\underline{Finite size scaling: Exact Vs Asymptotic predictions:}

\vspace{0.2cm}

As discussed in Sec.\ref{sec:universalformula}, the expression for the n'th Renyi entropy contains $n!$ terms, and only one of the them contributes to the volume law coefficient in the thermodynamic limit (compare Eq.\ref{eq:s3} and Eq.\ref{eq:sn} ). The asymptotic result, Eq.\ref{eq:sn}, also  matches with the Renyi entropies  $S^A_n(\overline{ \rho_A})$ (Eq.\ref{eq:avgrhosn}). It is worthwhile to compare these two predictions, the exact and the asymptotic, with exact diagonalization results. Fig.\ref{fig:finite_size_MBB_s3_av} compares the deviation of the exact MBB result for $S_3$ (Eq.\ref{eq:s3}) from the exact diagonalization, with the deviation of the asymptotic MBB result (Eq.\ref{eq:sn}) from the exact diagonalization. We notice that the exact result fares much better than the asymptotic one in a finite sized system. We also perform the finite-size scaling of the  deviation of the CTPQ state for $S_3$ from the exact diagonalization results. The extrapolation to thermodynamic limit indicates that the deviation becomes negative in the thermodynamic limit, which is again consistent with our prediction that $S_3$ for a chaotic system would be a convex function of $L_A$.

\subsection{Non-integrable Spin-$1/2$ Chain far from Integrability}
In this section, we consider the Hamiltonian given by Eq.\ref{eq:spin_model}
\[
H=\sum_i^L -Z_iZ_{i+1}-Z_i +X_i.
\]
Our goal is to compare the Renyi entropies obtained from the exact diagonalization of $H$ with our analytical predictions in Sec.\ref{sec:universalformula} based on the assumption that eigenstates behave as if they were chosen from the `ergodic bipartition' (EB) ensemble in Eq.\ref{eq:ergodic}.

We impose the periodic boundary condition $i\equiv i+L$ and choose $L=20$. The analytical prediction, say, for $S_2$ (Eq. \ref{eq:s2}) involves the knowledge of the density of states of $H_A$ and $H_{\overline{A}}$. One approximate way to proceed is $S^M_A(E_A) = s(E_A/V_A) L_A$ where $s(x)$ is the entropy density at energy density $x$ obtained from the largest size accessible within ED (here $L=20$). Alternatively, one can diagonalize $H_A$ and  $H_{\overline{A}}$ as well, and use the actual microcanonical density of states $S^M_A(E_A)$ and $S^M_{\overline{A}}(E-E_A)$  from such simulations. Here we chose this latter approach.

\begin{figure}
	\centering
	
	\begin{subfigure}[t]{0.5\textwidth}
		\includegraphics[width=\textwidth]{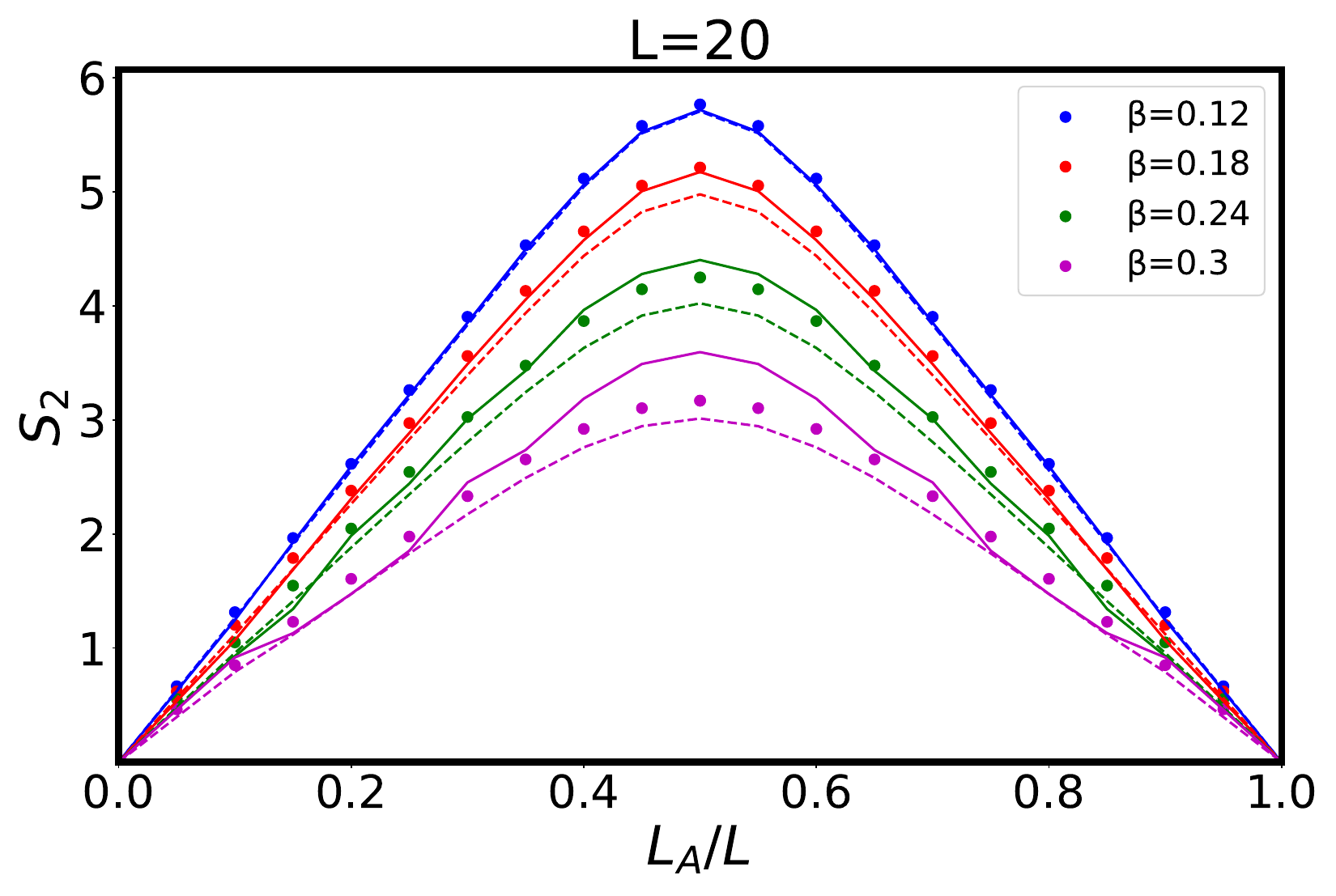}
		%\caption{$S_2$ given by eigenstates (solid dots), ergodic bipartition conjecture Eq.\ref{eq:s2} (solid lines) , CTPQ states Eq.\ref{eq:fujita} (dashed lines). }
	\end{subfigure}
    
\begin{subfigure}[t]{0.5\textwidth}
	\includegraphics[width=\textwidth]{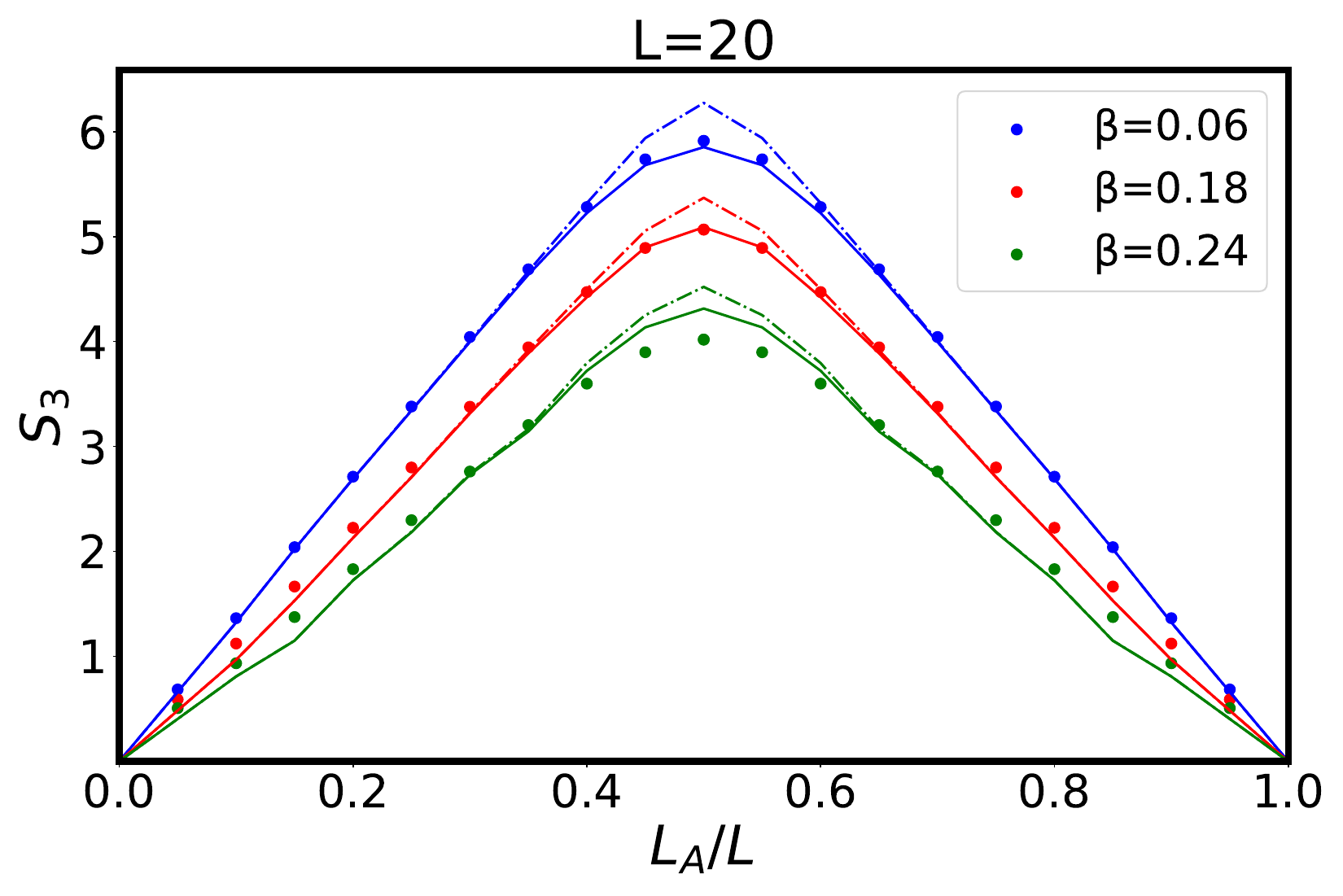}
		%\caption{$S_3$ given by eigenstates (solid dots), ergodic bipartition conjecture Eq.\ref{eq:s3} (solid lines) , ergodic bipartition conjecture with only taking the leading term Eq.\ref{eq:sn} (dashed lines). }
\end{subfigure}
	\caption{Comparison of the second Renyi entropy $S_2$ and the third Renyi entropy $S_3$ obtained by different methods for the eigenstates of 1D non-integrable Hamiltonian in Eq.\ref{eq:spin_model}. \textbf{Top figure:} Exact diagonalization (solid dots),    Ergodic bipartition states, Eq.\ref{eq:s2} (solid lines) and CTPQ states, Eq.\ref{eq:fujita} (dashed lines).
     \textbf{Bottom figure:} Exact diagonalization (solid dots),    Exact expression for ergodic bipartition states, Eq.\ref{eq:s3} (solid lines), Leading order expression for ergodic bipartition states, Eq.\ref{eq:sn} (dash-dot lines). }
	\label{fig:s2_spin_model_compare}
\end{figure}

Fig.\ref{fig:s2_spin_model_compare} compares our analytical prediction with ED. For $\beta \lesssim 0.3$, the predictions match rather well with the ED results. For smaller temperatures, there are slight deviations, which we attribute to the fact that the system sizes accessible within ED, the spectrum is not dense enough at the corresponding energy densities leading to a poor estimate of the density of states. We also show the comparison with CTPQ states. We notice that even at relatively high temperatures, $\beta \gtrsim 0.3$, the predictions from CTPQ do not fare well compared to those with the EB state.  

\begin{figure}
	\centering
	\begin{subfigure}[b]{0.45\textwidth}
		\includegraphics[width=\textwidth]{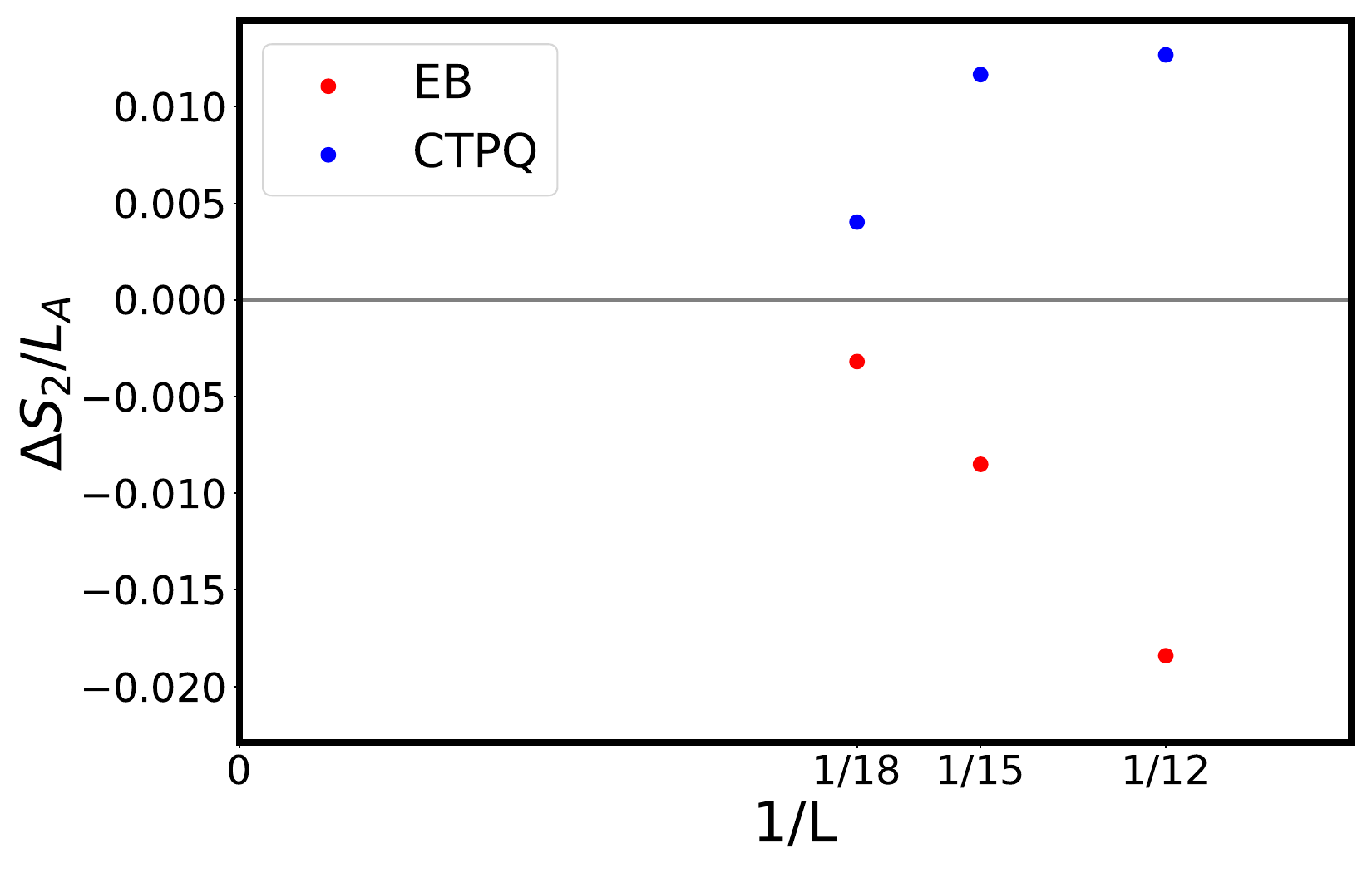}
	\end{subfigure}
	\begin{subfigure}[b]{0.45\textwidth}
		\includegraphics[width=\textwidth]{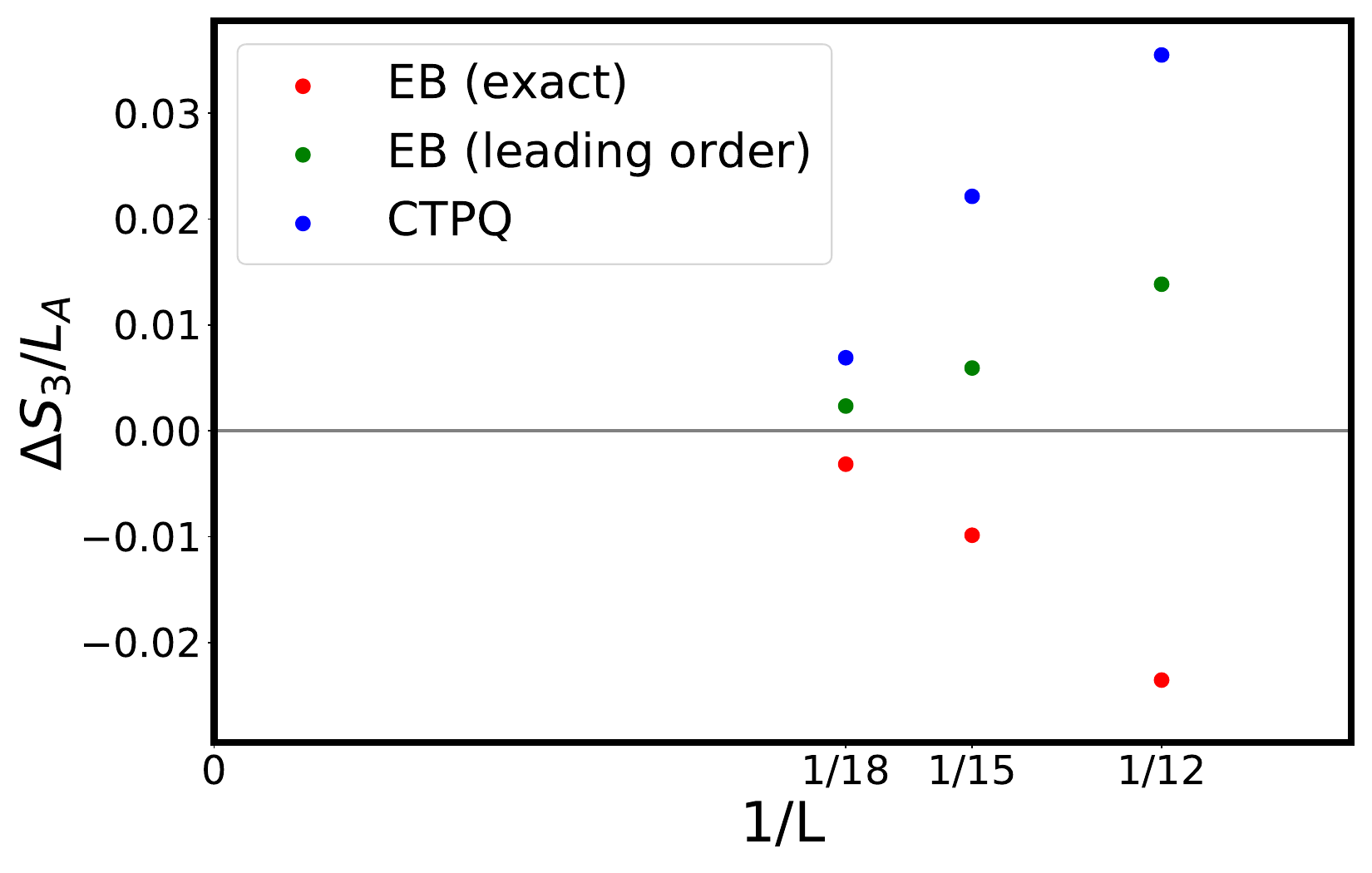}
	\end{subfigure}
	\begin{subfigure}[b]{0.45\textwidth}
		\includegraphics[width=\textwidth]{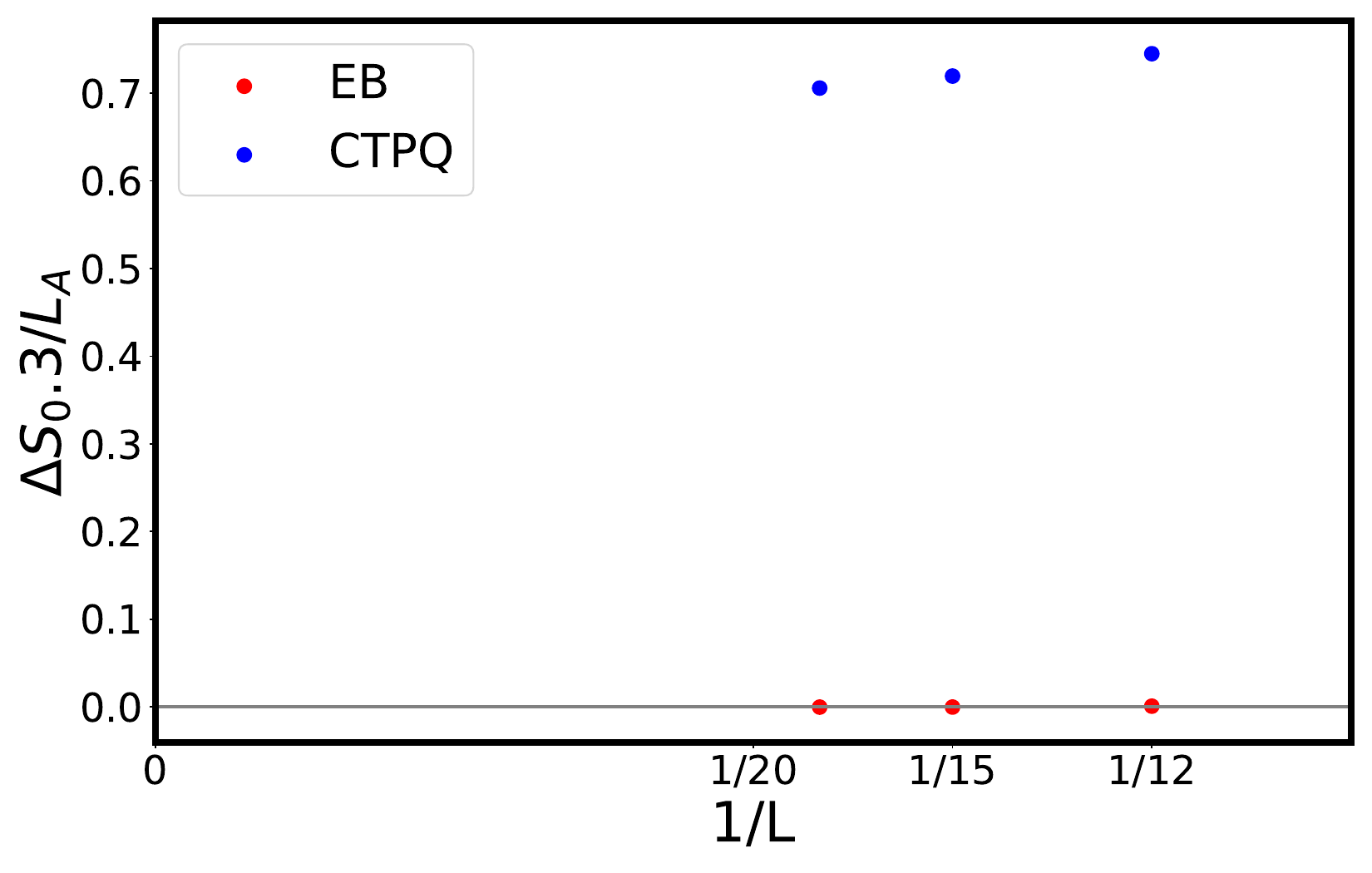}
	\end{subfigure}
	\caption{ Finite size scaling for of $\Delta S_n/L_A$ for ergodic bipartition (EB) states and CTPQ states where $\Delta S_n$ is defined as the difference between the analytical expressions for the corresponding states (EB or CTPQ) and the exact diagonalization results. The Hamiltonian is given by Eq.\ref{eq:spin_model} and eigenstates correspond to $\beta=0.06$.  $L_A$ is chosen to be $L/3$ for all $L$.  Note that in the middle panel,  ``EB (leading order)'' refers to the expression in Eq.\ref{eq:sn} while ``EB (exact)'' refers to the Eq.\ref{eq:s3}. In the top panel we use the exact expression for  EB states (Eq.\ref{eq:s2}) while in the bottom panel, we use the leading order result (Eq.\ref{eq:sn}) for  EB states.}
	\label{fig:finite_size_spin_model_epsilon_1}
\end{figure}

We also perform the finite size scaling for Renyi entropies, Fig.\ref{fig:finite_size_spin_model_epsilon_1} where $\Delta S_2$ denotes the deviation of the analytical prediction from the exact diagonalization results. The upper panel shows the finite size scaling for $S_2$ while the lower panel shows the results for $S_3$, where we also compare our asymptotic result (Eq.\ref{eq:sn}) with the more accurate result (Eq.\ref{eq:s3}). Similar to the case of MBB states in the previous section, we again find that EB states fare better compared to the CTPQ states, and for the EB states, the exact expression fares better than the asymptotic one.

\section{Summary and Discussion} \label{sec:discuss}

\begin{figure}
	\centering
	\includegraphics[width=0.5\textwidth]{{{Gaussian_s2_various_L_beta_0.6}.pdf}}
	\caption{Evolution of the shape of the Renyi entropy $S_2$  as the total system is increased for a system with Gaussian density of states  (Eq.\ref{eq:s2_gaussian}). Note that as $L \rightarrow \infty$, the Renyi entropy is a convex function for all $f$, and has a cusp singularity at $ f= 1/2$.} 
	\label{fig:Gaussian_s2_various_L_beta_0.6}
\end{figure}

\begin{figure}
	\centering
	\includegraphics[width=0.5\textwidth]{{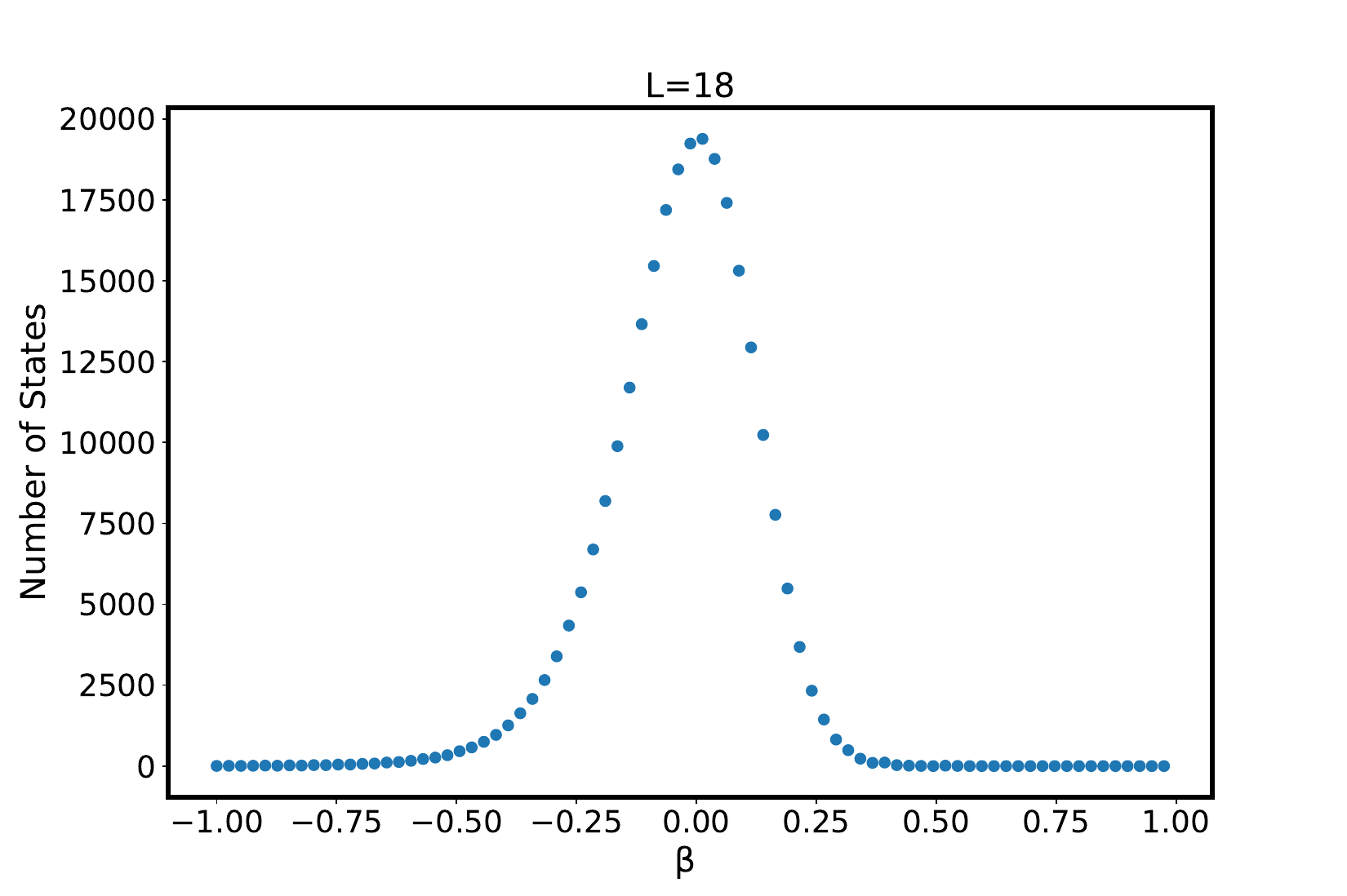}}
	\caption{The density of states as a function of  the temperature of the eigenstates  for a 18 site spin-chain. The temperature  for individual eigenstates $\ket{E_n}$ is evaluated by solving the equation $\frac{\tr \left(H e^{-\beta(E_n) H} \right)}{\tr \left( e^{- \beta(E_n) H}\right) } = E_n $. } 
	\label{fig:dos_beta}
\end{figure}

In this paper we derived a universal expression for the Renyi entropy of chaotic eigenstates for arbitrary subsystem to system ratio by employing arguments based on ergodicity. We found that Renyi entropy of chaotic eigenstates do not match the Renyi entropy of the corresponding thermal ensemble unless $ f = V_A/V$, the subsystem to total system ratio, approaches zero. For a general value of $f$, the Renyi entropy density $S_n/V_A$ has a non-trivial dependence on $f$, and only in the case of von Neumann entropy $n\rightarrow 1$, the density (i.e. the volume law coefficient) is independent of $f$. The curvature $\dfrac{d^2 S_n}{d f^2}$ is positive (negative) for $n>1$ ($n<1$) and therefore the volume law coefficient for $n > 1$ is greater (less) than that of a corresponding thermal ensemble. Such dependence is quite different than the Renyi entropies corresponding to (a) Thermal density matrix as well as CTPQ state \cite{sugiura2013, fujita2017universality}, for which the Renyi entropies densities are independent of $f$ (b) Free fermion systems for which the von Neumann entropies (and hence all Renyi entropies $n > 1$) are concave functions of $f$ \cite{singh2014, vidmar2017} (c) A random state in the Hilbert state \cite{lubkin1978,page1993average}, or systems without \textit{any} conservation laws \cite{husefloquet2015} for which all $S_n$ are simply given by $V f \log(2)$ ($f < 1/2$) and do not have any curvature dependence.   Our theoretical prediction matches rather well with the exact diagonalization results on quantum spin chains.

In exact diagonalization studies on finite systems, the  curvature dependence characteristic of the thermodynamic limit can be a bit challenging to observe.  In fact, most of the curvature seen in finite size systems can be attributed to the subleading terms in $S_n$ (e.g., the second term in the numerator of Eq.\ref{eq:s2}) which do not contribute to the volume law coefficient at any fixed $V_A/V$ in the thermodynamic limit. The presence of these terms in finite size systems can lead to the appearance that $S_n$ for $n > 1$ is a concave function of $V_A/V$ (see, e.g., Fig.\ref{fig:Gaussian_s2_various_L_beta_0.6}). Further, the magnitude of the curvature vanishes at infinite temperature, and is proportional to $\beta ^2$ at high temperatures. In exact diagonalization studies on finite systems, most states have $\abs{\beta}$ below $O(1)$ (see Fig.\ref{fig:dos_beta}), which also makes it harder to observe the curvature. 

Our result shows that  Renyi entropy for a given subsystem to total system volume fraction $f=V_A/V$ depends on the density of states at an energy density that is itself a function of $f$. This allows one to obtain information about the full spectrum of the Hamiltonian by keeping the Renyi index $n$ fixed and only varying $f$ from a single eigenstate. To demonstrate this, we expand the microcanonical entropy $s(u)$ at an energy density corresponding to the infinite temperature $T\to \infty$: $s(u)=\log 2+\alpha_2 u^2+\alpha_3u^3+\cdots$, where we choose $u=0$ corresponding to $T\to \infty$ without lose of generality. From Eq.\ref{eq:saddle}, one can solve for the saddle point energy density $u^*_{A}(\{\alpha_i \})$ and $u^*_{\overline{A}} (  \{\alpha_i \} )$, and plug them in Eq.\ref{eq:sn_saddle} to obtain an equation relating $S_n,f,\{\alpha_i\}$. Suppose that one can measure $S_n$ for various $f$ given an single eigenstate, we then have a system of equations of $\{\alpha_i\}$ given from different $(S_n,f)$. By solving these equations, one can construct the whole function $s(u)$ to have the full spectrum (density of state as a function of energy density) just from a single eigenstate. Note that this is in strong contrast to the limit $V_A/V \rightarrow 0$ where $S_n$ only encodes thermodynamical information at temperature $\beta^{-1}$ and $(n \beta)^{-1}$.

Our result also provides a particularly simple prediction for Renyi entropies of chaotic eigenstates for systems where the entropy density $s(u)$ depends on the energy density $u$ in a power law fashion i.e. $s(u) = c u^{\alpha}$ where $c$ is a constant. This is because in this case one can solve the saddle point equation (Eq.\ref{eq:saddle}) analytically. Consider, for example, a conformal field theory (CFT) in $d$ space dimensions, where the exponent $\alpha = \frac{d}{d+1}$. A straightforward calculation yields:

\begin{equation}
S_n = \frac{n}{(1-n)f} \left[ \{(1-f) + f n^{1/(\alpha-1) }\}^{1-\alpha} - 1 \right] S_1
%S_n = \frac{n}{1-n} c u^{\alpha} V \left[ \{(1-f) + f n^{1/(\alpha-1) }\}^{1-\alpha} - 1 \right]
\end{equation}
where the von Neumann entanglement entropy $S_1 = c u^{\alpha} V f$, i.e., it follows the Page curve as expected ($f < 1/2$ of course).

The dependence of Renyi entropies on subsystem to total system ratio sheds light on how to distinguish a mixed, thermal density matrix from a pure state which locally looks thermal. Besides being a basic question in quantum statistical mechanics, this question is also of central interest in `black hole information paradox' \cite{susskind_book, preskill1993}, where Hawking's calculation \cite{Hawking1975} implies that the radiation emanating from an evaporating black hole resembles a thermal system, while at the same time, if one were to describe the evaporation process by a unitary evolution of a pure quantum state, then one expects that there must exist correlations that distinguish the state of the black hole from a thermal state. Our results indicate that the dependence of Renyi entropy $S_n$ on $V_A/V$ may be one way to distinguish a thermal state from a pure state of a black hole.

In this paper, we focussed primarily on the volume law coefficient of the Renyi entropies. Ref.\cite{rigol_half2017} calculated the subleading contributions to the von Neumann entropy for the infinite temperature particle-number conserving states discussed in Ref.\cite{garrison2015does}, and put an upper bound that scales as $\sqrt{V}$ for $V_A/V = 1/2$. In similar spirit, it will be interesting to calculate the subleading contributions to the non-infinite temperature states introduced in this paper.

During the submission of this paper, we noticed that a recent work, Ref.\cite{huang2017},  also conjectures that states of the form Eq.\ref{eq:ergodic} may represent eigenstates of chaotic Hamiltonians. Ref. \cite{huang2017} argues that average of von Neumann entropy over \textit{all} eigenstates is linear in subsystem size at the leading order upto $V_A/V = 1/2$ with volume law coefficient $\log(2)$ for a spin-1/2 system. This is consistent with our results, and follows from our general formula, Eq.\ref{eq:sn}, for individual eigenstates: the average will be dominated by eigenstates at the infinite temperature, whose entanglement at the leading order is indeed $V_A \log(2)$ upto $V_A/V = 1/2$.

\vspace{0.2cm}

\textbf{\underline{Acknowledgements:}}   
We thank Ning Bao, Tom Faulkner, Hiroyuki Fujita, Jim Garrison, Tom Hartman, Jonathan Lam, John McGreevy, Yuya Nakagawa, Mark Srednicki, Sho Sugiura, Masataka Watanabe for  discussions, and especially to Jim Garrison and John McGreevy for comments on a draft. TG acknowledges support from the UCSD startup funds and is also supported as a Alfred P. Sloan
Research Fellow. This work used the Extreme Science and Engineering Discovery Environment (XSEDE) (see Ref.\cite{xsede}), which is supported by National Science Foundation grant number ACI-1548562. We thank KITP (Santa Barbara) where part of the manuscript was written while attending the workshop  ``Quantum Physics of Information''. This research was also supported in part by the National Science Foundation under Grant No. NSF PHY-1125915.  

%\input{v4.bbl}
%\bibliography{v1bib}
%\renewcommand\refname{Reference}
%\bibliographystyle{unsrt}

\onecolumngrid

%\section{Appendix}
\appendix
\section{Subsystem Energy Fluctuation in an Ergodic Bipartition(EB) state } \label{sec:energyfluct}

Consider an ergodic bipartition(EB) state defined by Eq.\ref{eq:ergodic}, 

\begin{equation}
\ket{E} =  \sum_{   E_i^A+E_j^{\overline{A}}  \in (E-\frac{1}{2}\Delta , E+\frac{1}{2} \Delta  )     }  C_{ij} \ket{ E_{i}^A} \otimes \ket{E_{j}^{\overline{A}}  },
\end{equation}

the probability of finding an eigenstate $\ket{E_i^A}$ on region $A$ is the diagonal element of reduced density matrix given by Eq.\ref{eq:rho_av}:

\begin{equation}
\bra{E_i^A} \rho_A \ket{E_i^A}=\frac{1}{N} e^{S_{\overline{A}}(E-E_i^A)},
\end{equation}

from which we can derive the probability of finding a state with energy $E_A$ by multiplying the density of state $e^{S_A(E_A)}$ on $A$:
\begin{equation}
P(E_A)\sim e^{S_A(E_A)}   e^{S_{\overline{A}}(E-E^A)}.
\end{equation}

This function has a peak at $E_A=\overline{E_A}$ determined by the saddle point equation 

\begin{equation}
\eval{\frac{\partial S_A(E_A)}{\partial E_A}}_{E_A=\overline{E_A}}  = \eval{\frac{\partial S_{\overline{A}}(E_{\overline{A}})}{\partial E_{\overline{A}}}  }_{E_{\overline{A}}=E-\overline{E_A}} .
\end{equation}

By expanding ${P(E_A)}$ around $\overline{E_A}$, $P(E_A)$ takes the Gaussian form:
\begin{equation}
P(E_A)\sim e^{-\frac{\left(E_A- \overline{E_A}  \right)^2}{2\Delta E^2}},
\end{equation}
with 
\begin{equation}
\Delta E^2 =cT^2\frac{V_AV_{\overline{A}}}{V_A+V_{\overline{A}}}=cT^2V f(1-f) 
\end{equation}
where $c$ denotes the specific heat per unit volume, $T$ denotes the temperature, and $f\equiv V_A/V$.

\section{Proof that $|S^A_n(\overline{\tr \rho^n_A}) - S^A_{n,\textrm{avg}}|$ is exponentially small in the total system size. } \label{sec:equiv}

Consider an ergodic bipartition (EB) ensemble defined by 
Eq.\ref{eq:ergodic}:

\begin{equation}
	\ket{E} =  \sum_{i,j}   C_{ij} \ket{ E_{i}^A} \otimes \ket{E_{j}^{\overline{A}}  },
\end{equation}
where $\{C_{ij}\}$ is chosen from the probability distribution function

\begin{equation}
	P(\{ C_{ij}  \}) \propto \delta (1-\sum_{ij}|C_{ij}|^2) \prod\limits_{i,j} \delta(E_i^A +E_j^A -E)
\end{equation}
where the the index $i (j) $ in $C_{ij}$ labels the state in $A (\overline{A})$. The reduced density matrix of $A$ can be obtained by tracing out the Hilbert space in $\overline{A}$:
\begin{equation} \label{eq:append_av_rho}
	\rho_A=\tr_{\overline{A}}\ket{\psi} \bra{\psi} =\sum_{i,i'}\ket{E_i^A}\bra{E_{i'}^A}  \sum_{j} C_{ij} C^*_{i'j}.
\end{equation}
In the main text we define two different averaging procedures for the Renyi entropy 

\begin{equation}\label{eq:append_sav}
	\begin{split}
		&S^A_n(\overline{\tr \rho^n_A}) = \frac{1}{1-n} \log\left( \overline{\tr \rho^n_A}\right)\\
		&S^A_{n,\textrm{avg}} =  \frac{1}{1-n} \overline{\log \left( \tr \rho^n_A \right)},
	\end{split}
\end{equation}
and state that the difference between these two vanishes in the volume of the system. Here we provide the proof for this claim.~\\

\noindent
First 
\begin{equation}\label{eq:trace}
	\tr \rho_A^n=\overline{\tr \rho_A^n} +\left( \tr \rho_A^n-  \overline{\tr \rho_A^n}    \right) =\overline{\tr \rho_A^n}  \left( 1+x   \right),
\end{equation}
where
\begin{equation}
	x\equiv \frac{\tr \rho_A^n}{\overline{\tr\rho^n_A}}-1 
\end{equation}
Plug Eq.\ref{eq:trace} into Eq.\ref{eq:append_sav}, we have 
\begin{equation}
	S^A_{n,\textrm{avg}} =  \frac{1}{1-n} \overline{\log \left( \tr \left(\rho^n_A \right)  \right)}=S^A_n(\overline{\tr \rho^n_A})  +\frac{1}{1-n} \overline{\log \left(1+x\right)  } ,
\end{equation}
where the last term, the difference between two averages, would be our main focus. By definition $\overline{x}=0$, and 
\begin{equation}\label{eq:x2}
	\overline{x^2} = \overline{\left(\frac{\tr \rho_A^n}{\overline{\tr\rho^n_A}}-1   \right)^2 } =\frac{\overline{\left( \tr \rho_A^n\right)^2 }}{\overline{\tr \rho_A^n}^2} -1.
\end{equation}
Via Eq.\ref{eq:append_av_rho}, 

\begin{equation}
	\rho_A^n=\sum_{i_1,j_1,k_1} \ket{E_{i_1}} \bra{E_{j_1}} C_{i_1k_1}C_{j_1k_1}^* 
	\sum_{i_2,j_2,k_2} \ket{E_{i_2}} \bra{E_{j_2}} C_{i_2k_2}C_{j_2k_2}^* 
	...
	\sum_{i_n,j_n,k_n} \ket{E_{i_n}} \bra{E_{j_n}} C_{i_nk_n}C_{j_nk_n}^* .
\end{equation}
By taking the trace of the above formula, we get

\begin{equation}
	\tr\rho_A^n =\sum_{i_1,j_1,k_1}  \sum_{i_2,j_2,k_2}... \sum_{i_n,j_n,k_n}  \delta_{j_1,i_2}\delta_{j_2,i_3}...\delta_{j_n,i_1}      C_{i_1k_1}C_{j_1k_1}^*C_{i_2k_2}C_{j_2k_2}^* ... C_{i_nk_n}C_{j_nk_n}^* .
\end{equation}
Now we are going to calculate the 2n point correlation function, which contains $n! $ terms:
\begin{equation}\label{eq:append_2n}
	\overline{C_{i_1k_1}C_{j_1k_1}^*C_{i_2k_2}C_{j_2k_2}^* ... C_{i_nk_n}C_{j_nk_n}^*} = \overline{C_{i_1k_1}C_{j_1k_1}^*} ...\overline{C_{i_nk_n}C_{j_nk_n}^*} +\text{All the other possible pairings.},
\end{equation}
Note that the above equality is only true when the dimension of the restricted Hilbert space $N\to \infty$ with $n$ being finite such that wick's theorem can hold. When we sum all the indices to calculate $\Tr\rho_A^n $, the term with the maximal number of summation for the state in $\mathcal{H_{\overline{A}}}$ (labelled by $k$) will exponentially dominates all the other terms. Looking back to Eq.\ref{eq:append_2n}, only first term contains no delta function constraint for $k$, and thus

\begin{equation}
	\begin{split}
		\overline{\tr\rho_A^n} &=\sum_{i_1,j_1,k_1}  \sum_{i_2,j_2,k_2}... \sum_{i_n,j_n,k_n}  \left[\delta_{j_1,i_2}\delta_{j_2,i_3}...\delta_{j_n,i_1}   \frac{1}{N^n} \delta_{i_1,j_1}...\delta_{i_n,j_n} + \cdots\right]\\ 
		&=\frac{1}{N^n}\sum_i \sum_{k_1,...k_n} \delta_{E_i^A+E_{k_1}^{\overline{A}},E} \quad ... \quad \delta_{E_i^A+E_{k_n}^{\overline{A}},E}  +\cdots \\
		&=\frac{\sum_{E_A}e^{S^M_A(E_A) +nS^M_{\overline{A}}(E-E_A)}}{  \left[\sum_{E_A}e^{S^M_A(E_A)+S^M_{\overline{A}}(E-E_A)}\right]^n} \left[ 1+ O(e^{-\alpha_1 V}) \right],
	\end{split}
\end{equation}
which gives

\begin{equation}\label{eq:av_den}
	\overline{\tr \rho_A^n}^2=   \left[\frac{   \sum_{E_A} e^{ S^M_A(E_A)+
			n S^M_{\overline{A}}(E-E_A) } }{ \left[\sum_{E_A}  e^{ S^M_A(E_A)+ S^M_{\overline{A}}(E-E_A) } \right]^{n}  } \right]^2 \left[ 1+O(e^{-\alpha_2 V})\right]
\end{equation}

with $\alpha_1$ and $\alpha_2$ are positive order 1 constants 
Similar calculation shows 
\begin{equation}\label{eq:av_num}
	\overline{\left( \tr \rho_A^n\right)^2} =   \left[\frac{   \sum_{E_A} e^{ S^M_A(E_A)+
			n S^M_{\overline{A}}(E-E_A) } }{ \left[\sum_{E_A}  e^{ S^M_A(E_A)+ S^M_{\overline{A}}(E-E_A) } \right]^{n}  } \right]^2 \left[ 1+O(e^{-\alpha_3 V})\right].
\end{equation}
Plug Eq.\ref{eq:av_num} and Eq.\ref{eq:av_den} into Eq.\ref{eq:x2} :
\begin{equation}
	\overline{x^2}=O(e^{-\alpha V}),
\end{equation}
where $\alpha$ is a positive order 1 constant. This means that in thermodynamic limit $V\to \infty$, there is no fluctuation of $x$, and the precise statement is given by 
\begin{equation}
	\text{Prob}\left( \left| x  \right| \geq \epsilon \right) \leq \frac{\overline{x^2}}{\epsilon^2}. 
\end{equation}
via Chebyshev's inequality, and thus there does not exist $x$ with finite distance away from zero in thermodynamics limit ($V \to \infty$), and a immediate consequence is that 

\begin{equation}
	\left| S^A_n(\overline{\tr \rho^n_A}) - S^A_{n,\textrm{avg}}    \right|  =\frac{1}{1-n} \overline{\log (1+x)} =O(e^{-\alpha V})
\end{equation} 
and thus the difference between these two averages decreases exponentially in volume.

\section{Second Renyi Entropy of an Ergodic Bipartition (EB) State}  \label{sec:S2calculation}

Here we provide the calculation of the averaged second Renyi entropy of a EB state. From a technical standpoint, the calculations are similar to those in Ref.\cite{fujita2017universality}. Consider an EB state in an energy window $I\equiv \left( E-\frac{1}{2}\Delta,E+\frac{1}{2}\Delta  \right)$
\begin{equation}
\ket{E} =\sum_{i,j} C_{ij} \ket{E_i^A, E_j^{\overline{A}}}
\end{equation}
where $\{C_{ij}\}$ is chosen from the probability distribution function
\begin{equation}
P(\{ C_{ij}  \}) \propto \delta (1-\sum_{ij}|C_{ij}|^2) \prod\limits_{i,j} \delta(E_i^A +E_j^A -E).
\end{equation}
Note that the the first index $i$ in $C_{ij}$ labels the state in $A$ while the second index $j$ labels the states in $\overline{A}$. Now we can calculate the reduced density matrix of $A$:
\begin{equation} \label{rhoA}
\rho_A=\tr_{\overline{A}}\ket{E} \bra{E} =\sum_{i,i'}\ket{E_i^A}\bra{E_{i'}^A}  \sum_{j} C_{ij} C^*_{i'j},
\end{equation}

and $\rho_A^2$ is

\begin{equation}
\rho_A^2
=\sum_{i,k'} \ket{E_i^A} \bra{E_{k'}^A} \sum_{i',j,l} C_{ij}C^*_{i'j} C_{i'l}C^*_{k'l}.
\end{equation}

Then it is straightforward to calculate $\tr \rho_A^2$ :
\begin{equation}
\tr \rho_A^2=\sum_{i,j,k,l} C_{ij}C_{kl}C^*_{il}C^*_{kj},
\end{equation}

In order to calculate the average of the second Renyi entropy:

\begin{equation}
\overline{S_2}=-{\log\left[  \overline{ \tr \rho_A^2} \right]},
\end{equation}

We perform the average for $  C_{ij}C_{kl}C^*_{il}C^*_{kj} $ first:
\begin{equation}
\overline{ C_{ij}C_{kl}C^*_{il}C^*_{kj}} =   \frac{1}{N(N+1)} \Big[ \delta_{jl} \delta_{E_i^A +E_j^{\overline{A}},E} \delta_{E_k^A+E_j^{\overline{A}},E}    +\delta_{ik} \delta_{E_i^A +E_j^{\overline{A}},E} \delta_{E_i^A +E_l^{\overline{A}},E}  \Big],
\end{equation}

where $N$ is the dimension of the Hilbert space in the restricted energy window.
Next we calculate $\overline{ \tr \rho_A^2}  $ :

\begin{equation} \label{rho2}
\begin{split}
\overline{ \tr \rho_A^2}&= \frac{1}{N(N+1)} \left[
\sum_{i,j,k} \delta_{E_i^A +E_j^{\overline{A}},E} \delta_{E_k^A +E_j^{\overline{A}},E}
+ \sum_{i,j,l} \delta_{E_i^A +E_j^{\overline{A}},E} \delta_{E_i^A +E_l^{\overline{A}},E} \right] \\
&=  \frac{1}{N(N+1)}  \left[
\sum_{E_A} e^{ 2S^M_A(E_A)+ S^M_{\overline{A}}(E-E_A) }
+ \sum_{E_{\overline{A}}} e^{ 2S^M_{\overline{A}}(E_{\overline{A}})  +  S^M_A(E-E_{\overline{A}}) } \right]\\
&=\frac{1}{N(N+1)}  \left[
\sum_{E_A} e^{ 2S^M_A(E_A)+ S^M_{\overline{A}}(E-E_A) }
+ e^{ S^M_A(E_A)  + 2 S^M_{\overline{A}}(E-E_{\overline{A}}) } \right],
\end{split}
\end{equation}
where we make the change of variable for the last term. Note that the above equation is manifestly symmetric between $A$ and $\bar{A}$. Finally we can derive the second Renyi entropy of an EB state:
\begin{equation}
\begin{split}
\overline{S_2}&=- \log \left[\frac{1}{N^2}   \left[ \sum_{E_A} e^{ 2S^M_A(E_A)+ S^M_{\overline{A}}(E-E_A) }
+ e^{ S^M_A(E_A)+   2S^M_{\overline{A}}(E-E_A) } \right]   \right]\\
& = - \log \left[\frac{   \sum_{E_A} e^{ 2S^M_A(E_A)+ S^M_{\overline{A}}(E-E_A) }+ e^{ S^M_A(E_A)+
		2S^M_{\overline{A}}(E-E_A) } }{   \left[\sum_{E_A}   e^{ S^M_A(E_A)+ S^M_{\overline{A}}(E-E_A) } \right]^2} \right],
\end{split}
\end{equation}
where we have assumed $N$ is large such that $N+1\approx N$.
Notice that when we take $V_A,V \to \infty $ with $\frac{V_A}{V} <\frac{1}{2}$, the first term in the numerator can be neglected, and thus
\begin{equation} \label{S2A}
\overline{S_2}= - \log \left[\frac{   \sum_{E_A} e^{ S^M_A(E_A)+
		2S^M_{\overline{A}}(E-E_A) } }{ \left[\sum_{E_A}  e^{ S^M_A(E_A)+ S^M_{\overline{A}}(E-E_A) } \right]^2  } \right].
\end{equation}
We will show below that this is exactly the second Renyi entropy of the reduced density matrix of $A$ obtained from maximally mixed state.

\section{Renyi Entropy $S_n$ of an Ergodic Bipartition (EB) state}  \label{sec:Sncalculation}

Eq.\ref{rhoA} shows the reduced density matrix obtained from a EB state:
\begin{equation}
\rho_A =\sum_{i,j,k} \ket{E_i} \bra{E_j} C_{ik}C_{jk}^* ,
\end{equation}
where as usual the first index of $C$ label the eigenstate in $\mathcal{H}_A$ and the second index of $C$ labels the eigenstate in $\mathcal{H}_{\overline{A}}$.

Next we can calculate $\rho_A ^n$:

\begin{equation}
\rho_A^n=\sum_{i_1,j_1,k_1} \ket{E_{i_1}} \bra{E_{j_1}} C_{i_1k_1}C_{j_1k_1}^* 
\sum_{i_2,j_2,k_2} \ket{E_{i_2}} \bra{E_{j_2}} C_{i_2k_2}C_{j_2k_2}^* 
...
\sum_{i_n,j_n,k_n} \ket{E_{i_n}} \bra{E_{j_n}} C_{i_nk_n}C_{j_nk_n}^* .
\end{equation}

By taking the trace of the above formula, we get

\begin{equation}
\tr\rho_A^n =\sum_{i_1,j_1,k_1}  \sum_{i_2,j_2,k_2}... \sum_{i_n,j_n,k_n}  \delta_{j_1,i_2}\delta_{j_2,i_3}...\delta_{j_n,i_1}      C_{i_1k_1}C_{j_1k_1}^*C_{i_2k_2}C_{j_2k_2}^* ... C_{i_nk_n}C_{j_nk_n}^* .
\end{equation}

Now we are going to calculate the 2n point correlation function, which contains $n! $ terms:
\begin{equation}\label{eq:2npoint}
\overline{C_{i_1k_1}C_{j_1k_1}^*C_{i_2k_2}C_{j_2k_2}^* ... C_{i_nk_n}C_{j_nk_n}^*} = \overline{C_{i_1k_1}C_{j_1k_1}^*} ...\overline{C_{i_nk_n}C_{j_nk_n}^*} +\text{All the other possible pairings.},
\end{equation}
Note that the above equality is only true when the dimension of the restricted Hilbert space $N\to \infty$ with $n$ being finite such that wick's theorem can hold. When we sum all the indices to calculate $\tr\rho_A^n $, the term with the maximal number of summation for the state in $\mathcal{H_{\overline{A}}}$ (labelled by $k$) will exponentially dominates all the other terms. Looking back to Eq.\ref{eq:2npoint}, only first term contains no delta function constraint for $k$, and thus

\begin{equation}
\begin{split}
\overline{\tr\rho_A^n} &=\sum_{i_1,j_1,k_1}  \sum_{i_2,j_2,k_2}... \sum_{i_n,j_n,k_n}  \delta_{j_1,i_2}\delta_{j_2,i_3}...\delta_{j_n,i_1}   \frac{1}{N^n} \delta_{i_1,j_1}...\delta_{i_n,j_n}\\ 
&=\frac{1}{N^n}\sum_i \sum_{k_1,...k_n} \delta_{E_i^A+E_{k_1}^{\overline{A}},E} \quad ... \quad \delta_{E_i^A+E_{k_n}^{\overline{A}},E} \\
&=\frac{1}{N^n}\sum_{E_A}e^{S^M_A(E_A) +nS^M_{\overline{A}}(E-E_A)}.
\end{split}
\end{equation}

Finally we can obtain the Renyi entropy of order $n$ in thermodynamic limit:
\begin{equation}
\overline{S_{n}}= \frac{1}{1-n} \log \left[\frac{   \sum_{E_A} e^{ S^M_A(E_A)+
		n S^M_{\overline{A}}(E-E_A) } }{ \left[\sum_{E_A}  e^{ S^M_A(E_A)+ S^M_{\overline{A}}(E-E_A) } \right]^{n}  } \right].
\end{equation}
which is exactly equal to the Renyi entropy obtained from the maximally mixed state.
In general we can derive the closed form of the Renyi entropy for arbitrary order without taking thermodynamic limit by calculating Eq.\ref{eq:2npoint} explicitly, but for simplicity, we only present the exact result for $n=3$:
\begin{equation}
\overline{S_3}=-\frac{1}{2}  \log \left[  \frac{   \sum_{E_A} e^{ S^M_A(E_A)+
		3 S^M_{\overline{A}}(E-E_A) }  + 3e^{ 2S^M_A(E_A)+
		2S^M_{\overline{A}}(E-E_A) }  +e^{ S^M_A(E_A)+
		S^M_{\overline{A}}(E-E_A) }  +e^{ 3S^M_A(E_A)+
		S^M_{\overline{A}}(E-E_A) }   }{ \left[\sum_{E_A}  e^{ S^M_A(E_A)+ S^M_{\overline{A}}(E-E_A) } \right]^{3}  }              \right]
\end{equation}

\section{Curvature of Renyi entropy $S_n$ } \label{sec:saddle}

\begin{figure}
	\centering
	\includegraphics[width=0.5\textwidth]{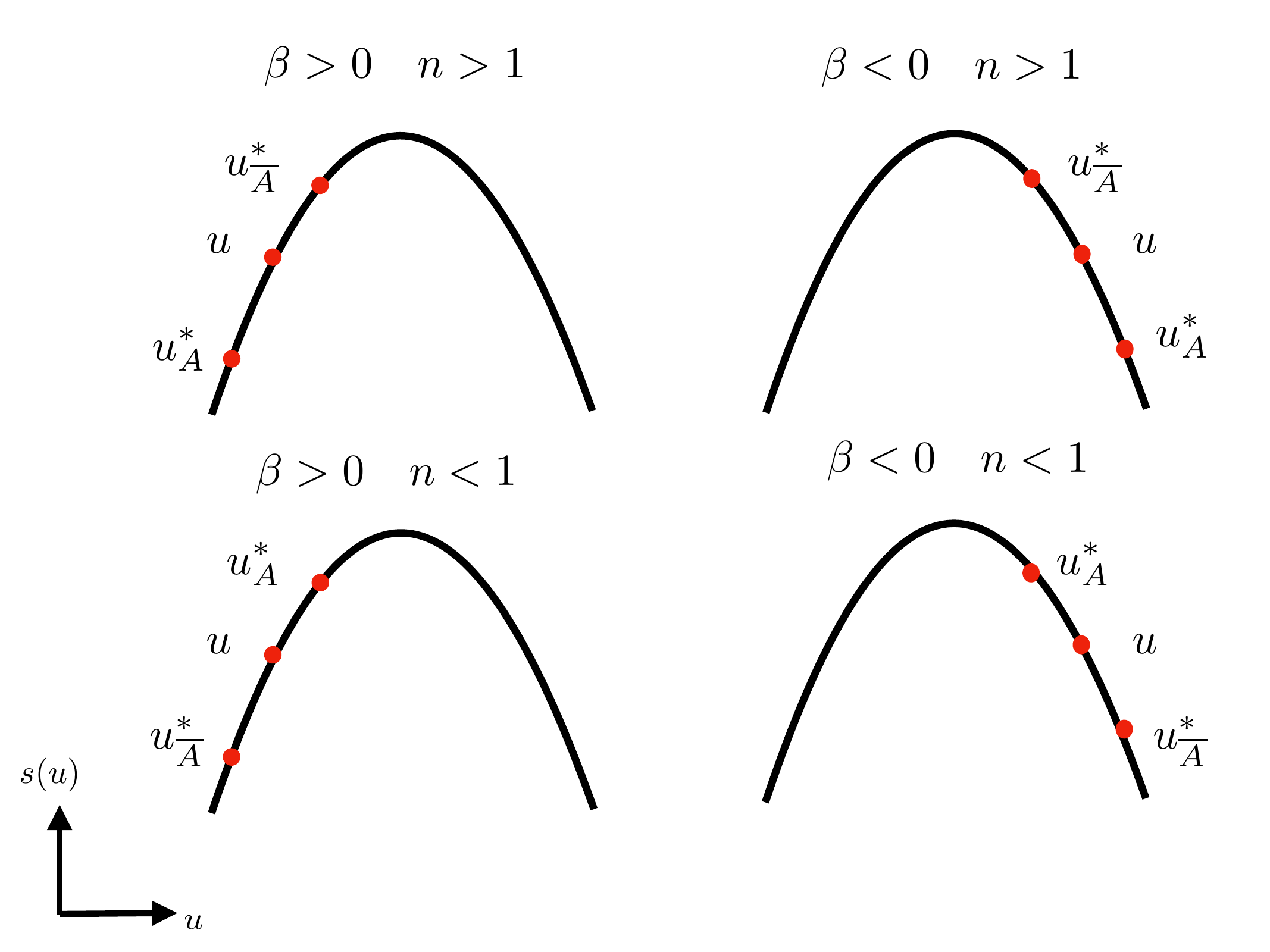}
	\caption{Allowed relative positions of the energies $u, u^{*}_A$ and $u^*_{\overline{A}}$ that solve the Eq. \ref{eq:saddle}. Note that the concavity of $s(u)$ curve, imposed by the non-negative value of specific heat, plays a crucial role.}
	\label{fig:saddle}
\end{figure}

Here we show the Renyi entropy  $\overline{S_n}$ is convex for $n>1$ while concave for $n<1$. Recall that  $\overline{S_n}$ is given by Eq.\ref{eq:sn_saddle}
\begin{equation}\label{eq:append_sn}
\overline{S_n}=\frac{V}{1-n}\left[ fs(u^{*}_A)+n(1-f)s(u^*_{\overline{A}}) - n s(u)  \right].
\end{equation}

By taking the derivative of Eq.\ref{eq:append_sn}, we have 
\begin{equation}\label{eq:s'}
\frac{1-n}{V}\frac{\partial \overline{S_n}}{\partial f}=s(u^{*}_A)+f\frac{\partial s(u^{*}_A)}{\partial u^{*}_A}\frac{\partial u^{*}_A}{\partial f} -ns(u^*_{\overline{A}}) +n(1-f)\frac{\partial s(u^*_{\overline{A}})}{\partial u^*_{\overline{A}}} \frac{\partial u^*_{\overline{A}}}{\partial f} .
\end{equation}

With the saddle point equation Eq.\ref{eq:saddle}
\begin{equation}\label{eq:append_saddle}
\frac{\partial s(u^{*}_A)}{\partial u^{*}_A}=n\frac{\partial s(u^*_{\overline{A}})}{\partial u^*_{\overline{A}}}
\end{equation}
and
\begin{equation}\label{eq:uAbar}
\frac{\partial u^*_{\overline{A}}}{\partial f}=\frac{1}{1-f} \left[ u^*_{\overline{A}}-u^{*}_A - f\frac{\partial u^{*}_A}{\partial f}\right] 
\end{equation}

obtained by differentiating the energy conservation condition  $fu^{*}_A+(1-f)u^*_{\overline{A}}=u$, 

Eq.\ref{eq:s'} can be simplified as 

\begin{equation}\label{eq:s'_simplified}
\frac{(1-n)}{V}\frac{\partial \overline{S_n}}{\partial f}=s(u^{*}_A)-ns(u^*_{\overline{A}}) +\frac{\partial s(u^{*}_A)}{\partial u^{*}_A}(u^*_{\overline{A}}-u^{*}_A). 
\end{equation}

Now we differentiate Eq.\ref{eq:s'_simplified} with respect to $f$ again:

\begin{equation}\label{eq:s''}
\frac{(1-n)}{V}\frac{\partial^2 \overline{S_n}}{\partial f^2}=
\frac{\partial s(u^{*}_A)}{\partial u^{*}_A} \frac{\partial u^{*}_A}{\partial f }-n\frac{\partial s(u^*_{\overline{A}})}{\partial u^*_{\overline{A}}} \frac{\partial u^*_{\overline{A}}}{\partial f }+\frac{\partial ^2s(u^{*}_A)}{\partial u^{*}_{A^2}} \frac{\partial u^{*}_A}{\partial f} (u^*_{\overline{A}} -u^{*}_A)  +\frac{\partial s(u^{*}_A)}{\partial u^{*}_A} \left(  \frac{\partial u^*_{\overline{A}}}{\partial f} -\frac{\partial u^{*}_A}{\partial f} \right).
\end{equation}

With Eq.\ref{eq:append_saddle} and Eq.\ref{eq:uAbar}, Eq.\ref{eq:s''} can be simplified:

\begin{equation}\label{eq:curvature}
\frac{(1-n)}{V}\frac{\partial^2 \overline{S_n}}{\partial f^2}=\frac{\partial^2 s(u^{*}_A)}{ \partial u^{*}_{A^2}} \frac{\partial u^{*}_A}{\partial f} (u^*_{\overline{A}}-u^{*}_A  ).
\end{equation}

Now let's study the sign of the R.H.S. The first quantity $s''(u^{*}_A)$ is always negative due to the concavity of microcanonical entropy. The sign of the last quantity $u^*_{\overline{A}} - u^{*}_A$ can also be shown via the concavity of microcanonical entropy and the saddle point equation Eq.\ref{eq:append_saddle}, 

\begin{equation}\label{eq:sign_1}
\text{Sgn} (u^*_{\overline{A}} - u^{*}_A)=
\begin{cases}
\text{Sgn}(n-1) & \text{for} \quad \beta>0\\
-\text{Sgn}(n-1)  &\text{for} \quad \beta<0,
\end{cases}
\end{equation}

where $\beta \equiv \frac{\partial s(u)}{\partial u}  $. See Fig.\ref{fig:saddle} for a graphical illustration.

As for the sign of the quantity in the middle $\frac{\partial u^{*}_A}{\partial f}$, we need to differentiate the saddle point equation Eq.\ref{eq:append_saddle} with respect to $f$:
\begin{equation}
\frac{\partial ^2s(u^{*}_A)}{\partial u^{*}_{A^2}}\frac{\partial u^{*}_A}{\partial f} =n \frac{\partial ^2s(u^*_{\overline{A}})}{\partial u_{\overline{A}^2}}\frac{\partial u^*_{\overline{A}}}{\partial f}. 
\end{equation} 
This implies $\frac{\partial u^{*}_A}{\partial f} $ and $ \frac{\partial u^*_{\overline{A}}}{\partial f}$ have the same sign. Combining this fact with the energy conservation condition Eq.\ref{eq:uAbar}, we have

\begin{equation}\label{eq:sign_2}
\text{Sgn} \left(\frac{\partial u^{*}_A}{\partial f}\right)=
\begin{cases}
\text{Sgn}(n-1) & \text{for} \quad \beta>0\\
-\text{Sgn}(n-1)  &\text{for} \quad \beta<0.
\end{cases}
\end{equation}

Finally by combining Eq.\ref{eq:curvature}, Eq.\ref{eq:sign_1}, Eq.\ref{eq:sign_2}, and the concavity of the microcanonical entropy, we obtain the final result

\begin{equation}
\begin{split}
&\frac{\partial^2 \overline{S_n}}{\partial f^2}>0  \quad  \text{for} \quad  n>1\\
&\frac{\partial^2 \overline{S_n}}{\partial f^2}<0  \quad  \text{for} \quad n<1 
\end{split}
\end{equation}
for all $\beta$ and $f = V_A/V$.

\section{Renyi entropy for a system with Gaussian density of states}\label{sec:Gaussian_derivation}

\centerline{ \textbf{   \underline{ Second Renyi Entropy  }   }}~\\

Suppose that the probability density of finding a state with energy $E$ takes the form:

\begin{equation}
P(E)=\frac{1}{\sqrt{2\pi V}}e^{-\frac{E^2}{2V}},
\end{equation}
we can then derive the density of state by multiplying the total number of states in the Hilbert space:

\begin{equation}\label{dos}
D(E)\sim 2^V e^{-\frac{E^2}{2V}}=e^{V\left[\log2 -\frac{1}{2}(\frac{E}{V})^2\right]}, 
\end{equation}
which implies the microcanonical entropy density $s$ is

\begin{equation}\label{micro_s}
s=\log2 -\frac{1}{2}u^2
\end{equation}
with $u$ denoting the energy density. Also we can define the inverse temperature $\beta$

\begin{equation}
\beta=\frac{\partial s}{\partial u} =-u.
\end{equation}

Given Eq.\ref{dos} or Eq.\ref{micro_s}, we can then calculate the number of states in $A$ and $\overline{A}$ with energy $E_A$ and $E_{\overline{A}}$ respectively:

\begin{equation}
e^{S_A(E_A)}=2^{V_A}P_A(E_A)\Delta=2^{V_A} \frac{1}{\sqrt{2\pi V_A}}e^{-\frac{E_A^2}{2V_A}}\Delta,
\end{equation}

\begin{equation}
e^{S_{\overline{A}}(E-E_A)}=2^{V_{\overline{A}}}P_{\overline{A}}(E_{\overline{A}})\Delta=2^{V_{\overline{A}}} \frac{1}{\sqrt{2\pi V_{\overline{A}}}}e^{-\frac{E_{\overline{A}}^2}{2V_{\overline{A} }}} \Delta,
\end{equation}
where $\Delta $ is the width of the energy window.
First we calculate 
\begin{equation}\label{denominator}
\begin{split}
\sum\limits_{E_ A}   e^{S_A(E_A)}e^{S_{\overline{A}}(E-E_A)}&=\frac{2^{V_A+V_{\overline{A}}}}{\sqrt{4\pi^2V_AV_{\overline{A}}}} \sum_{E_A}\Delta^2 e^{-\frac{E_A^2}{2V_A} - \frac{(E-E_A)^2}{2V_{\overline{A}}}} \\
&=\frac{2^{V_A+V_{\overline{A}}}}{\sqrt{\pi^2V_AV_{\overline{A}}}}  \Delta e^{-\frac{E^2}{V}}\sqrt{\frac{\pi}{\frac{1}{2V_A}+\frac{1}{2V_{\overline{A}}}}}\\
&=\frac{2^{V}}{\sqrt{2\pi V}}  \Delta e^{-\frac{E^2}{2V}},
\end{split}
\end{equation}
where we approximate $\sum_{E_A} \Delta$ by the continuous integral $\int dE_A$ and evaluate the Gaussian integral in the expression. 

The other quantity we need to evaluate is

\begin{equation}\label{left}
\begin{split}
\sum\limits_{E_ A}   e^{S_A(E_A)}e^{2S_{\overline{A}}(E-E_A)}&=\frac{2^{V_A+2V_{\overline{A}}}}{\sqrt{8\pi^{3}V_AV^2_{\overline{A}}}} \sum_{E_A}\Delta^3 e^{-\frac{E_A^2}{2V_A} - \frac{(E-E_A)^2}{V_{\overline{A}}}} \\
&=\frac{2^{V_A+2V_{\overline{A}}}}{2\pi\sqrt{(2V_A+V_{\overline{A}})V_{\overline{A}}}}  \Delta^2 e^{-\frac{E^2}{2V_A+V_{\overline{A}}}},
\end{split}
\end{equation}
and we also have

\begin{equation}\label{right}
\begin{split}
\sum\limits_{E_ A}   e^{2S_A(E_A)}e^{S_{\overline{A}}(E-E_A)}&=\frac{2^{2V_A+V_{\overline{A}}}}{\sqrt{8\pi^{3}V_A^2V_{\overline{A}}}} \sum_{E_A}\Delta^3 e^{-\frac{E_A^2}{V_A} - \frac{(E-E_A)^2}{2V_{\overline{A}}}} \\
&=\frac{2^{2V_A+V_{\overline{A}}}}{2\pi\sqrt{(V_A+2V_{\overline{A}})V_{A}}}  \Delta^2 e^{-\frac{E^2}{V_A+2V_{\overline{A}}}},
\end{split}
\end{equation}

With Eq.\ref{denominator}, Eq.\ref{left}, and Eq.\ref{right}, we can get the second renyi entropy :
\begin{equation}
\begin{split}
\overline{S_2}&= - \log \left[\frac{   \sum_{E_A} e^{ S^M_A(E_A)+ 2S^M_{\overline{A}}(E-E_A) }+ e^{ 2S^M_A(E_A)+S^M_{\overline{A}}(E-E_A) } }{   \left[\sum_{E_A}   e^{ S^M_A(E_A)+ S^M_{\overline{A}}(E-E_A) } \right]^2} \right]\\
&=-\log\left[\frac{1}{\sqrt{1-f^2}}e^{-V\gamma(f,u)} +\frac{1}{\sqrt{1-(1-f)^2}}e^{-V\gamma(1-f,u)} \right]
\end{split}
\end{equation}
where $u\equiv \frac{E}{V}$, $f\equiv \frac{V_A}{V}$, and 

\begin{equation}
\gamma (f,u)=f\log 2 -\frac{f}{1+f}u^2
\end{equation}

Notice that the $S_2$ is manifestly invariant under $f \to 1-f $, and is universal in the sense that it only depends on the energy density, and is capable of capturing the finite size correction of entanglement Renyi entropy. 

In thermodynamic limit $V\to \infty $ with $f<1/2$, we can then get 
\begin{equation}
\overline{S_2}=fV\left[  \log2-\frac{u^2}{1+f}   \right]=fV\left[  \log2-\frac{\beta^2}{1+f}   \right]
\end{equation}~\\

\centerline{ \textbf{ Renyi Entropy $S_n$ in the limit $V\to \infty $}   }~\\

In thermodynamic limit $V\to \infty$, we can solve for the saddle point equation 
\begin{equation}
\frac{\partial s(u_A^*)}{\partial u_A^*} =n \frac{\partial s(u^*_{\overline{A}})}{\partial u^*_{\overline{A}}}
\end{equation}
with $u^*_{\overline{A}} =\frac{u^*}{1-f}-\frac{f}{1-f}u^*_A$, and then plug it in to Eq.\ref{eq:sn_saddle}

\[
\overline{S_n}=\frac{V}{1-n}\left[ fs(u_A^*)  +n(1-f)s(u^*_{\overline{A}}) -ns(u)  \right]
\]
to derive $n$-th Renyi entropy.

First from the saddle point equation, we get 
\begin{equation}
u^*_A=nu^*_{\overline{A}}=n\left(\frac{u}{1-f}-\frac{f}{1-f}u^*_A \right),
\end{equation}
from which we can solve for $u^*_A$:
\begin{equation}
u^*_A=\frac{nu}{1+(n-1)f}.
\end{equation}

Finally we can then calculate Renyi entropy for arbitrary Renyi index $n$:

\begin{equation}
\begin{split}
\overline{S_n}&=\frac{V}{1-n}\left[ fs(u_A^*)  +n(1-f)s(u^*_{\overline{A}}) -ns(u)  \right]\\
&=\frac{V}{1-n}\left[  f  \left(  \log 2-\frac{1}{2} (u_A^*)^2   \right)     +n(1-f)\left( \log2-\frac{(u_A^*)^2}{2n^2}   \right)   -n\left( \log2-\frac{u^2}{2}\right)  \right]\\
&=fV\left[\log2-\frac{nu^2}{  2 (1+(n-1)f)}\right]\\
&=fV\left[\log2-\frac{n\beta^2}{2(1+(n-1)f )}\right].
\end{split}
\end{equation}

When $n\to 1$, we have 
\begin{equation}
\frac{\overline{S_1}}{fV}= \log2 -\frac{1}{2}\beta^2,
\end{equation}
which is exactly the microcanonical entropy density.

\section{Some Mathematical Results on Correlation Functions for Random Vectors }  \label{sec:mathcorr}

%\centerline{ \textbf{   \underline{Correlation Function for a Random Vector in $\mathbb{R}^M$}   }}~\\

Suppose that we have a random vector $\bm{X}$ in $\mathbb{R^M}$ with the probability distribution function being
\begin{equation}
P( \{ x_i \} ) \propto \delta( 1-\sum_{i=1}^M x_i^2),
\end{equation}
where $  \{ x_i \} $ denotes all the component of $\bm{X}$. Note the probability measure is invariant under $O(M)$, which immediately indicates that
\begin{equation}
\left\langle x_ix_j  \right\rangle =0  \quad \forall i\neq  j.
\end{equation}
For the case where $i=j$, we recall the constraint:
\begin{equation}
\sum_{i=1}^M x_i^2 =1.
\end{equation}

When we take average for the equation above, due to the $O(M)$ symmetry,  $\left\langle x_i^2 \right\rangle =  \left\langle x_j^2 \right\rangle \quad \forall i,j$, and thus we can get
\begin{equation}
\left\langle x_i^2 \right\rangle =\frac{1}{M}.
\end{equation}

As for the four point function $\left\langle  x_ix_jx_kx_l\right\rangle$, by imposing the $O(M)$ symmetry, we can write down the most general form:
\begin{equation}
\left\langle  x_ix_jx_kx_l\right\rangle= A \left[\delta_{ij}\delta_{kl} +\delta_{ik}\delta_{jl}+\delta_{il}\delta_{kj}    \right].
\end{equation}
Now in order to determine $A$, we contract the indices $k,l$ first, meaning we set $k=l$ and then perform summation over $k$:
\begin{equation}
\left\langle  x_ix_j\right\rangle= A \sum_{k=1}^M \left[\delta_{ij}  +\delta_{ik}\delta_{jk}+\delta_{ik}\delta_{kj}    \right]
=A\delta_{ij}\left[M+1+1   \right] .
\end{equation}

Recall that $\left\langle  x_ix_j  \right\rangle  =\frac{1}{M}\delta_{ij}$, and thus $A$ can be determined:
\begin{equation}
A=\frac{1}{M(M+2)},
\end{equation}
meaning the four point function is
\begin{equation}\label{eq:fourpoint}
\left\langle  x_ix_j x_k x_l\right\rangle=  \frac{1}{M(M+2)} \left[\delta_{ij}\delta_{kl} +\delta_{ik}\delta_{jl}+\delta_{il}\delta_{kj}    \right].
\end{equation}

Notice that Eq.\ref{eq:fourpoint} looks very similar to Wick's theorem, but actually it is not:
\begin{equation}
\begin{split}
\left\langle  x_ix_j x_k x_l\right\rangle&=  \frac{1}{M(M+2)} \left[\delta_{ij}\delta_{kl} +\delta_{ik}\delta_{jl}+\delta_{il}\delta_{kj}    \right]  \\
&\neq \frac{1}{M^2} \left[\delta_{ij}\delta_{kl} +\delta_{ik}\delta_{jl}+\delta_{il}\delta_{kj}   \right]  \\
&=\left\langle x_ix_j\right\rangle \left\langle x_kx_l\right\rangle +\left\langle x_ix_k\right\rangle \left\langle x_jx_l\right\rangle  +\left\langle x_ix_l\right\rangle \left\langle x_kx_j\right\rangle.
\end{split}
\end{equation}

However we can notice that when we take $M\to \infty $, the difference between these two approaches zero! This is not a coincidence since when we randomly pick a vector from $\mathbb{R}^M$ with the only constraint being the magnitude of the vector and $M$ is large, we can show that the  probability distribution function for $ \{ x_i |i=1,2,...s \}$ is Gaussian for $s\ll M$:
\begin{equation}
\begin{split}
P(x_1,x_2,...x_s)& = \prod\limits_{i=s+1}^M\int_{-\infty} ^{\infty} dx_i P(x_1,x_2,...x_M)\\
&=  \prod\limits_{i=s+1}^M \int_{-\infty} ^{\infty} dx_i   \delta( 1-\sum_{i=1}^s x_i^2-  \sum_{i=s+1}^M x_i^2)\\
&\propto \left[ 1-\sum_{i=1}^s x_i^2 \right]^{\frac{M-s-1}{2}},
\end{split}
\end{equation}
where we used fact that the $M-s$ dimensional integral is proportional to the surface area of  $M-s$ dimensional ball with radius $R=\sqrt{1-\sum_{i=1}^s x_i^2}$. Then

\begin{equation}
P(x_1,x_2,...x_s)\sim \left[ 1-\sum_{i=1}^s x_i^2 \right]^{\frac{N-s-1}{2}}
\sim  \left[ 1-\frac{1}{N\sigma^2}\sum_{i=1}^s x_i^2 \right]^{\frac{N}{2}}
\sim e^{-\frac{\sum_{i=1}^s x_i^2   }{2\sigma^2}},
\end{equation}
where the variance $\sigma^2=\frac{1}{N}$. Therefore, the probability distribution function the small number of degrees of freedom is indeed a Gaussian! Also note that the derivation above is just the standard derivation from microcanonical ensemble to canonical ensemble. For example, consider $M$ particles in a box with total energy being $E$, in microcanonical ensemble we can write down the probability distribution for momenta $\{ p_i \}$:
\begin{equation}
P( \{ p_i \} ) \propto \delta( E-\sum_{i=1}^M \frac{p_i^2}{2m}),
\end{equation}
where we consider kinetic energy only for simplicity. Then if we look at the probability function for small numbers of particles, we can derive the Boltzmann distribution for those particles via the exactly the same calculation above, which is indeed a Gaussian in momenta.~\\

\centerline{\textbf{   \underline{Correlation Functions for a Random State without Imposing any Constraint}}}~\\

Given a Hilbert space $\mathcal{H}=\mathcal{H_A} \otimes \mathcal{H_{\overline{A}}}$ with Dim($\mathcal{H}$)$\equiv N$, suppose we pick a state :
\begin{equation}
\ket{\psi} =\sum_{i,j} C_{ij} \ket{E_i^A, E_j^{\overline{A}}},
\end{equation}

where $\{C_{ij}\}$ is chosen from the probability distribution function
\begin{equation}
P(\{ C_{ij}  \}) \propto \delta (1-\sum_{ij}|C_{ij}|^2)
\end{equation}
with $i=1,2,...\text{Dim}(\mathcal{H_A}) $ and  $i=1,2,...\text{Dim}(\mathcal{H_{\overline{A}}})$ respectively. Since $C_{ij} =u_{ij}+iv_{ij}$,
a random pure state is equivalent to a vector in $ \mathbb{R}^{M}  \quad( M=2N) $ with the length of the vector being one, meaning it can be regarded as a point on $S^{M-1}$ with the probability measure:

\begin{equation}
P(\{ u_{ij}\}, \{ v_{ij}  \}) \propto \delta (1-\sum_{ij}u_{ij}^2 -\sum_{ij}v_{ij}^2)  .
\end{equation}

We may want to calculate the two point function:

\begin{equation}
\begin{split}
\left\langle  C_{ij} C_{kl} \right\rangle &=\left\langle  (u_{ij}+iv_{ij})( u_{kl}+iv_{kl} ) \right\rangle\\
&= \left\langle u_{ij} u_{kl}   \right\rangle  -   \left\langle v_{ij} v_{kl}   \right\rangle
+i \left\langle u_{ij} v_{kl}   \right\rangle  +   i \left\langle v_{ij} u_{kl}   \right\rangle \\
&= \left\langle u_{ij} u_{kl}   \right\rangle  - \left\langle v_{ij} v_{kl}   \right\rangle  ,
\end{split}
\end{equation}
where the last two terms vanish since $u_{ij}$ and $v_{kl}$ are different component $\forall i,j,k,l $ of a vector in $\mathbb{R}^M$. On the other hand,
\begin{equation}
\left\langle u_{ij} u_{kl}   \right\rangle = \left\langle u_{ij} u_{kl}   \right\rangle =\frac{1}{M}\delta_{ik} \delta_{jl},
\end{equation}
and thus we conclude
\begin{equation}
\left\langle  C_{ij} C_{kl} \right\rangle =0  \quad \forall i,j,k,l .
\end{equation}

Let's consider another two point function $  \left\langle  C_{ij} C^*_{kl} \right\rangle   $:

\begin{equation}
\begin{split}
\left\langle  C_{ij} C^*_{kl} \right\rangle&  = \left\langle  (u_{ij}+iv_{ij})( u_{kl}-iv_{kl} ) \right\rangle\\
&= \left\langle u_{ij} u_{kl}   \right\rangle  + \left\langle v_{ij} v_{kl}   \right\rangle  \\
&=\frac{2}{M}\delta_{ik} \delta_{jl}\\
&=\frac{1}{N}\delta_{ik} \delta_{jl}.
\end{split}
\end{equation}
Note that the above result can also be recognized as
\begin{equation}
\left\langle  C_{ij} C^*_{kl} \right\rangle  =\delta_{kj}\delta_{jl}  \left\langle |C_{ij}|^2   \right\rangle= \frac{1}{N}\delta_{ik}  \delta_{jl}
\end{equation}
The lesson here is that $C_{ij}$ is only correlated with its conjugate counterpart. \\

We can also consider the four point function:
\begin{equation}
\left\langle  C_{ij} C_{kl} C^*_{mn} C^*_{pq} \right\rangle = \left\langle  (u_{ij}+iv_{ij})  (u_{kl}+iv_{kl})   (u_{mn}-iv_{mn}) (u_{pq}-iv_{pq}) \right\rangle
\end{equation}
There are $16$ terms in the expansion, but the terms with odd number of $u$ vanish. Thus,

\begin{equation}
\begin{split}
\left\langle  C_{ij} C_{kl} C^*_{mn} C^*_{pq} \right\rangle &=     \left\langle u_{ij}u_{kl}u_{mn}u_{pq} \right\rangle \\
&-\left\langle u_{ij}u_{kl}v_{mn}v_{pq} \right\rangle +\left\langle u_{ij}u_{mn}v_{kl}v_{pq} \right\rangle +
\left\langle u_{ij}u_{pq}v_{kl}v_{mn} \right\rangle \\
&+\left\langle u_{kl}u_{mn}v_{ij}v_{pq} \right\rangle +\left\langle u_{kl}u_{pq}v_{ij}v_{mn} \right\rangle -
\left\langle u_{mn}u_{pq}v_{ij}v_{kl} \right\rangle \\
&+\left\langle v_{ij}v_{kl}v_{mn}v_{pq} \right\rangle ,
\end{split}
\end{equation}
where the first line and the last line correspond to the term with four and zero number of $u$, and the $C^4_2$ terms in between are from choosing two $u$ and two $v$.
Via Eq.\ref{eq:fourpoint}, the first and the last term are
\begin{equation} \label{eq:14}
\begin{split}
\left\langle u_{ij}u_{kl}u_{mn}u_{pq} \right\rangle&=\left\langle v_{ij}v_{kl}v_{mn}v_{pq} \right\rangle \\
&= \frac{1}{M(M+2)} \left[ \delta_{ik} \delta_{jl} \delta_{mp} \delta_{nq} +  \delta_{im} \delta_{jn} \delta_{kp} \delta_{lq}    +\delta_{ip} \delta_{jq} \delta_{km} \delta_{ln}\right],
\end{split}
\end{equation}

while six terms in the middle are

\begin{equation}\label{eq:middle}
\begin{split}
&-\left\langle u_{ij}u_{kl}v_{mn}v_{pq} \right\rangle +\left\langle u_{ij}u_{mn}v_{kl}v_{pq} \right\rangle +
\left\langle u_{ij}u_{pq}v_{kl}v_{mn} \right\rangle \\
&+\left\langle u_{kl}u_{mn}v_{ij}v_{pq} \right\rangle +\left\langle u_{kl}u_{pq}v_{ij}v_{mn} \right\rangle -
\left\langle u_{mn}u_{pq}v_{ij}v_{kl} \right\rangle\\
&=\frac{2}{M(M+2)} \left[ \delta_{ik} \delta_{jl} \delta_{mp} \delta_{nq} + \delta_{im} \delta_{jn} \delta_{kp} \delta_{lq}   + \delta_{ip} \delta_{jq} \delta_{km} \delta_{ln}\right]
\end{split}
\end{equation}

Combining the result of Eq.\ref{eq:14} and Eq.\ref{eq:middle}, the four point function can be calculated :
\begin{equation}
\begin{split}
&\left\langle  C_{ij} C_{kl} C^*_{mn} C^*_{pq} \right\rangle \\
&=\frac{4}{M(M+2)} \left[ \delta_{im} \delta_{jn} \delta_{kp} \delta_{lq} + \delta_{ip} \delta_{jq} \delta_{km} \delta_{ln}\right]\\
&=\frac{1}{N(N+1)} \left[  \delta_{im} \delta_{jn} \delta_{kp} \delta_{lq} +  \delta_{ip} \delta_{jq} \delta_{km} \delta_{ln}\right]
\end{split}
\end{equation}

To check this result, we can calculate $\tr \rho_A^2$ without energy constraint, and we get back to the same answer in Ref\cite{lubkin1978}:
\begin{equation}
\overline{\tr\rho_A^2} = \frac{ \text{Dim}(\mathcal{H_A}) +  \text{Dim}(\mathcal{H_{\overline{A}}})}{ \text{Dim}(\mathcal{H_A})\text{Dim}(\mathcal{H_{\overline{A}}})+1}.
\end{equation}

From this result we can calculate the second Renyi entropy
\begin{equation}
\overline{S_2}=-\log \overline{\tr \rho_A^2}=  \log  \left[\text{Dim}(\mathcal{H_A}) \right]
\end{equation}
when we take both $\text{Dim}(\mathcal{H_A}) $ and $\text{Dim}(\mathcal{H_{\overline{A}}})$ to infinity while the ratio $ \text{Dim}(\mathcal{H_A})/\text{Dim}(\mathcal{H_{\overline{A}}})<1$.

Via Jensen's inequality, we have
\begin{equation}
\overline{S_2} = \log \left[\text{Dim}(\mathcal{H_A})  \right]\leq \overline{S_1} \leq  \log \left[ \text{Dim}(\mathcal{H_A})\right]
\end{equation}

and thus the entanglement entropy $\overline{S_1}$ is also maximal:

\begin{equation}
\overline{S_1}  = \log \left[\text{Dim}(\mathcal{H_A}) \right],
\end{equation}
which is the answer from Page's calculation \cite{page1993average}. ~\\

\centerline{\textbf{  \underline{Correlation Functions for a Random State at a Fixed Energy}}}	~\\

Consider a pure state in a small energy window with energy $E$:
\begin{equation}
\ket{\psi} =\sum_{i,j} C_{ij} \ket{E_i^A, E_j^{\overline{A}}}
\end{equation}
where $\{C_{ij}\}$ is chosen from the probability distribution function
\begin{equation}
P(\{ C_{ij}  \}) \propto \delta (1-\sum_{ij}|C_{ij}|^2) \prod\limits_{i,j} \delta(E_i^A +E_j^A -E),
\end{equation}
with $i=1,2,...\text{Dim}(\mathcal{H_A}) $ and  $i=1,2,...\text{Dim}(\mathcal{H_{\overline{A}}})$ respectively. Due to the energy conservation, the two point function will be
\begin{equation}
\left\langle  C_{ij} C^*_{kl} \right\rangle=  \frac{1}{N}\delta_{ik} \delta_{jl}\delta_{E_i^A +E_j^{\overline{A}} ,E},
\end{equation}
and the four point function is
\begin{equation}
\left\langle  C_{ij} C_{kl} C^*_{mn} C^*_{pq} \right\rangle =\frac{1}{N(N+1)} \Big[
\delta_{im} \delta_{jn} \delta_{kp} \delta_{lq}\delta_{E_i^A +E_j^{\overline{A}} ,E}  \delta_{E_k^A +E_l^{\overline{A}} ,E} 
+ \delta_{ip} \delta_{jq} \delta_{km} \delta_{ln} \delta_{E_i^A +E_j^{\overline{A}} ,E} \delta_{E_k^A +E_l^{\overline{A}} ,E}  \Big].
\end{equation}

 \end{document}